\documentclass{article}

\usepackage{comment}
\usepackage{ulem}
\usepackage{cancel}
\usepackage{subcaption} 

\usepackage{float}
\usepackage{graphicx} 
\usepackage{cite}
\usepackage{booktabs}
\usepackage{amsmath}
\usepackage{amssymb}
\usepackage{graphicx}
\usepackage{esint}
\usepackage{hyperref}
\usepackage{multirow}

\usepackage{color}

\title{A complex logistic equation for universal energy evolution in hadronic elastic scattering}

\author{Anderson Kendi Kohara}

\date{    $^1$CPHT, CNRS, École Polytechnique, Institut Polytechnique de Paris, 91120 Palaiseau, France \footnote{email:anderson.kendi@gmail.com}
\\%
}

\begin{document}

\maketitle

\begin{abstract}

We introduce a universal evolution equation for elastic scattering of hadrons, derived from Regge Field Theory (RFT) and solved in closed analytical form. The equation has a complex logistic structure and evolves initial amplitude profiles from existing models at a fixed energy, reproducing both differential cross sections and integrated quantities over a broad energy range. It admits a unique solution for each initial condition and rigorously satisfies unitarity, the Froissart--Martin bound, and dispersion relations. The dynamics are governed by two physically meaningful parameters: the effective Pomeron mass $\epsilon_{\mathcal{P}}$ and the nonlinear coupling $\lambda$, both fitted at a single energy.  By decoupling the nonperturbative input from the universal energy evolution, the framework enables model-independent extrapolations and provides a minimal predictive alternative to eikonal resummation. Moreover, the structure of the equation -- featuring rapidity evolution, saturation, and impact-parameter dependence -- shares qualitative features with nonlinear QCD equations at small-$x$, such as BK and JIMWLK, suggesting a possible bridge between Regge-based and QCD-based approaches to high-energy scattering.

\end{abstract}

\section{Introduction}
Understanding hadronic interactions at high energies remains one of the central challenges of strong-interaction physics. While perturbative QCD provides a systematic framework for processes with a hard scale, elastic scattering lies firmly in the nonperturbative regime. In this context, effective descriptions such as Regge theory \cite{Collins:1977jy,Gribov:1968fc}  and the eikonal formalism have played a key role in modeling total and differential cross sections. However, these approaches often rely on phenomenological inputs and face limitations in satisfying key theoretical constraints such as unitarity and the Froissart--Martin bound \cite{Froissart:1961,Martin:346321} at asymptotic energies.

In this work, we propose an alternative approach: we formulate high-energy elastic scattering as a nonlinear evolution equation in energy, grounded in Regge Field Theory (RFT). Our starting point is the observation that traditional single-Pomeron exchange models \cite{Donnachie:1992ny,Block:2006hy} lead to amplitudes that grow too rapidly with energy, violating unitarity and failing to capture saturation effects observed at the LHC and beyond. While eikonalization is a common remedy, it introduces ambiguities and model dependence in the treatment of multiple scatterings. 

Instead, we propose a minimal nonlinear generalization of the linear Regge evolution equation, featuring a complex logistic structure. The equation admits a closed-form solution and describes how an input amplitude profile in impact parameter space evolves with energy. The dynamics are controlled by only two physically meaningful parameters: an effective Pomeron intercept $\epsilon_P$, and a nonlinear coupling $\lambda$, related to triple-Pomeron interactions. This simplicity contrasts with more elaborate models that require fitting many energy-dependent parameters. Importantly, the evolution respects the key theoretical principles:
\begin{enumerate}
\item     Unitarity, ensured by the saturation of the imaginary part of the amplitude;

\item     Froissart-Martin bound, satisfied for a wide class of initial profiles;

\item     Analyticity, confirmed via dispersion relations and Cauchy-Riemann conditions;

\item     Uniqueness, guaranteed by Lipschitz continuity of the evolution flow.
\end{enumerate}

A central feature of the formalism is the decoupling of model dependence from energy evolution: the initial amplitude profile $T(s_0, b)$ can be taken from any phenomenological model that fits data at a reference energy $s_0$ (e.g., KFK \cite{kohara:2014}, BSW\cite{BSW:1979
}, DL\cite{Donnachie:2013xia
}, RealBB\cite{Bialas:2006kw}), while the subsequent evolution in energy is universal and governed solely by the nonlinear equation with only two dynamical parameters, $\epsilon_P$ and $\lambda$. This enables model-independent extrapolations to higher energies --- from ISR to LHC --- without refitting or adding arbitrary terms.

The use of existing phenomenological models for $T(s_0,b)$ is not a limitation but a deliberate choice to test the universality of the evolution: once the profile is fixed at one energy, the subsequent energy dependence is a genuine prediction of the nonlinear equation. In this sense, the nonperturbative transverse structure is supplied by the initial profile, while the evolution law captures the universal saturation dynamics. 

This separation of roles between the input profile and the evolution law is directly analogous to the treatment of parton distribution functions (PDFs) in perturbative QCD: PDFs are fitted to experimental data at an initial scale $Q_0^2$ and then evolved to other scales via universal, process-independent evolution equations such as DGLAP\cite{DGLAP} or BK\cite{Balitsky:1995ub,Kovchegov:1999yj,Marquet:2005qu}. In our case, the impact-parameter profile $T(s_0,b)$ plays the role of the PDF, and our nonlinear RFT-inspired equation is the analog of the evolution operator. Once the initial condition is fixed at one energy, the energy dependence of elastic observables is a genuine prediction of the formalism.

Moreover, the evolution equation shares structural similarities with nonlinear diffusion-type equations that appear in the context of QCD small-$x$ physics, such as the Balitsky–Kovchegov (BK) and JIMWLK \cite{Balitsky:1995ub,JalilianMarian:1997jx,JalilianMarian:1997gr,JalilianMarian:1997dw,Kovner:2000pt,Kovner:1999bj,Weigert:2000gi,Iancu:2000hn,Ferreiro:2001qy} equations. While our focus is on soft, nonperturbative scattering, this analogy points to a deeper connection between saturation dynamics and the evolution of elastic amplitudes.

 In our framework  we focus on elastic scattering of
identical hadrons (e.g. pp or p\=p) at high energy. The expressions
derived in the present work are intended to apply to forward elastic scattering dominated
by vacuum quantum number exchange, for which the impact-parameter
representation provides a natural and transparent description.

This paper is organized as follows: In Section 2, we derive the nonlinear evolution equation from phenomenological arguments, and obtain a closed-form solution in the diffusionless limit. Section 3 is devoted to the mathematical properties of the solution, including unitarity, analyticity, uniqueness of solution and asymptotic bounds. In section 4 we implement our equation and present numerical results using data-driven models as initial conditions, demonstrating the power and generality of the formalism. We conclude in Section 5 with a summary and future perspectives. In appendix A we present the kinematical factors, the definitions of the amplitudes and the connection with the experimental quantities, in appendix B we derive our evolution equation from Regge field theory and in appendix C we present useful integrals.

\section{ An evolution equation for elastic scattering}

In problems with spherical symmetry, such as scattering, it is well known that the amplitude can be expressed through a partial wave expansion. For axially symmetric cases, the scattering amplitude takes the form
\begin{eqnarray}
    \mathcal{A}(\theta)=16\pi\sum_{\ell=0}^{\infty}\,(2\ell+1)\frac{e^{i\,\delta_{\ell}}}{k}\sin \delta_\ell\,P_\ell(\cos\theta) ~.
\end{eqnarray}
Here, $\theta$ denotes the scattering angle and $\delta_\ell$ represents the phase shift. Geometrically, the angle of $\theta$ is related to the impact parameter between the two colliding objects, while the phase shift $\delta_\ell$ encodes the effect of the interaction potential between the colliding particles. Although commonly used in quantum mechanics, this decomposition is a general result that applies to wave phenomena governed by spherical symmetry, including classical wave scattering.

In relativistic scattering,  the Mandelstam variables $s$, $t$, and $u$  are used to describe the kinematics. Moreover, the scattering amplitudes must be symmetrized to account for the exchange symmetries of the system. The partial wave expansion can be analytically continued into the $t$-channel and reformulated as \cite{donnachie_dosch_landshoff_nachtmann_2002}
\begin{eqnarray}
    \mathcal{A}^{\pm}(s,t)=16\pi\sum_{\ell=0}^{\infty}\,(2l+1)A_\ell(t)\,\Big(1\pm e^{-i\pi\,\ell}\Big)\,P_\ell(\cos\theta_t) ~, 
\end{eqnarray}
with $\cos(\theta_t)=1+\frac{2s}{t-4m^2}$ and $A_\ell(t)$ carrying the phase shift information. The sign $\pm$  refers to even and odd angular momentum in the sum. In high-energy limit the Legendre polynomial approximates to $P_\ell(\cos(\theta_t))\sim s^\ell$. The analytic continuation of $\ell$ is necessary in order to have a convergent series. This is the well-known Regge formalism and the angular momentum is replaced by the Regge trajectory \cite{Chew_Regge_1962}
\begin{eqnarray}
    \ell=\alpha(t)=\alpha_0+\alpha'\,t~~,
\end{eqnarray}
describing the position of the poles in complex $\ell$-plane.
For center-of-mass energies from 1 to 20 GeV, the collision energy is considered high enough to apply Regge phenomenology but not sufficiently high to be classified as a hadronic high-energy process, since resonances and bound states could still contribute to the total cross section. For the elastic scattering experiments performed within this range, the Regge phenomenology  describes relatively well the behavior of total cross sections and the differential cross section near the origin, where the scattering angle is close to zero. The coefficients of the Regge trajectories found by the phenomenology give negative intercepts, leading to a decreasing total cross section with increasing energy, since in this framework the total cross section scales as $\sigma_t \sim s^{\alpha_0-1}$. Typically the Regge amplitude is written as 
\begin{eqnarray}
    \mathcal{A}^{\pm}(s,t)\sim 16\pi\sum_{k}\beta_k^{\pm}(t)\,\,\left(1\pm e^{-i\pi\alpha_k^{\pm}(t)}\right)\,\left(\frac{s}{s_0}\right)^{\alpha_k^{\pm}(t)} ~, 
\end{eqnarray}
where the sum runs over the specific family of Reggeons, $\beta_k^{\pm}(t)$ is the residue function describing how strongly the Regge pole contributes to the scattering amplitude and $\eta_k^{\pm}=(1\pm e^{-i\pi\alpha_k^{\pm}(t)})$ is the signature factor.

By the end of the 1960s, the prevailing view was that the total cross section should either saturate to a constant or vanish asymptotically \cite{Kinoshita:1970}. This view was supported by both  experimental trends and the Pomeranchuk theorems. However, the Froissart-Martin bound \cite{Froissart:1961, Martin:346321}, establishing an upper bound for the total cross section at asymptotic energies, gave an indication that an increasing cross section with increasing energy would be acceptable, at least if the strong interactions were considered a short-range interaction and the quantum probability of all possible scattering channels was conserved. Some experts, like Cheng and Wu, based on field theory of high-energy quantum electrodynamics, summing ladder diagrams predicted that the total cross section should increase with increasing energy \cite{Cheng:1970bi}, paving the way to the Pomeron interpretation.
 Three years latter, the  experiments conducted by Tevatron at FERMILAB and ISR at CERN \cite{Carroll:1974yv, Amati:1975} were able to access energies above 20 GeV and showed that the total cross section $\sigma_{\rm tot.}$ was indeed increasing. To explain the experimental data, the phenomenology required a different kind of Regge trajectory, the Pomeron. This object has  even parity across the channels pp and p\=p, its Regge trajectory contains a positive intercept and a smaller slope compared to the standard Regge trajectories. The simple-pole Pomeron amplitude is a complex function
\begin{eqnarray}
    \mathcal{A}_{\mathcal{P}}(s,t) = \,g_a(t)\,g_b(t)\,\beta_{\mathcal{P}}(t)\left(i-\cot\frac{\pi\alpha_{\mathcal{P}}(t)}{2}\right)\,\left(\frac{s}{s_0}\right)^{\alpha_{\mathcal{P}}(t)}  ~,\label{eq:pomeron-amplitude0}
\end{eqnarray}
where $\alpha_{\mathcal{P}}(t)=1+\epsilon_{\mathcal{P}}+\alpha^\prime\,t~$ is the standard linear Pomeron trajectory with the intercept $\epsilon_{\mathcal{P}}$ and the slope $\alpha'$ and positive parity $\eta_{\mathcal{P}}^{+}=-\sin(\pi\alpha_{\mathcal{P}})\left[i-\cot(\frac{\pi\alpha_{\mathcal{P}}}{2})\right]$. The functions $g_i(t)$ are the hadronic form factors and $\beta_{\mathcal{P}}(t)$ is the Pomeron residue function, which gives the coupling of the Pomeron with the external particles. 

Despite the initial success, traditional soft Pomeron amplitudes, which grow as a power of energy and lack transverse dynamics, face serious limitations at LHC and asymptotic energies. They violate unitarity constraints and as a consequence, the Froissart bound; fail to account for multiple scattering effects; and cannot describe the impact-parameter structure or saturation phenomena observed experimentally. These shortcomings motivate the search for alternative formulations that incorporate nonlinear dynamics, unitarization mechanisms, and an explicit transverse space treatment.

To better understand the high-energy behavior of elastic scattering amplitudes in impact parameter space ($b$-space), we analyze a simplified form of the soft Pomeron exchange.  The phenomenological values of $\epsilon_{\mathcal{P}}$ and $\alpha'$ (in the GeV$^{-2}$ scale) are typically small, and since we are interested in forward scattering where 
$t$ is also small, it is reasonable to expand the signature factor in powers of 
$\epsilon_{\mathcal{P}}+\alpha'\,t$. In particular,\footnote{Noting that $\cot\Big(\frac{\pi} {2}\alpha_{\mathcal{P}}(t)\Big)= -\tan\Big(\frac{\pi}{2}(\epsilon_{\mathcal{P}}+\alpha'\,t)\Big)$} we approximate
\begin{eqnarray}
    \cot\frac{\pi\alpha_{\mathcal{P}}(t)}{2}\sim -\frac{\pi}{2}(\epsilon_{\mathcal{P}}+\alpha'\,t)~.
\end{eqnarray}
This simplifies the form of the amplitudes, avoiding trigonometric functions. 
Assuming the typical form factors and the residue function as pure exponential forms
\begin{eqnarray}
    g_k(t)=A_k\,e^{\beta_k\,t} ~~~k=\{a,b\}~ ,
\end{eqnarray}
and
\begin{eqnarray}
    \beta_{\mathcal{P}}(t)=\beta_0\,e^{\beta_{\mathcal{P}}\,t}~, 
\end{eqnarray}
then, Eq.(\ref{eq:pomeron-amplitude0}) can be analytically Fourier transformed to the impact parameter space:
\begin{eqnarray}
    \label{eq:pomeron-amplitude}
   \widetilde{T}(s,b) &=&\frac{4\pi}{s}\int \frac{d^2q_{\perp}}{2\pi}\,e^{-i\,\vec{q}_{\perp}\cdot \vec{b}_{\perp}}\, \mathcal{A}(s,-q^2 ) \\ \nonumber
     &=&  \bar{A}\,s^{\epsilon_{\mathcal{P}}}\,\frac{e^{-b^2/4\mathcal{B}(s)}}{2\,\mathcal{B}(s)}\,\left[i+\frac{\pi}{2}\left(\epsilon_{\mathcal{P}}+\alpha'\frac{(b^2-4\,\mathcal{B}(s))}{4\,\mathcal{B}(s)^2}\right)\right]~~,
\end{eqnarray}
where $\bar{A}=4\pi\,A_a\,A_b\,\beta_{0}$ and $\mathcal{B}(s)=(\beta_a+\beta_b+\beta_{\mathcal{P}}+\alpha'\log s)$.  The transformation to impact-parameter space in Eq.~(\ref{eq:pomeron-amplitude}) is performed 
in the center-of-mass frame, where the beam axis defines the longitudinal 
direction and $b$ is the transverse distance between the projectile and 
target trajectories. This representation is not manifestly Lorentz covariant, 
since it singles out a preferred frame, but it is physically meaningful 
in the high-energy (Regge) limit, where longitudinal and transverse 
degrees of freedom are kinematically well separated. It has long been 
used in Regge theory, eikonal models, and Glauber formalisms. This transformation by itself does not affect the fundamental analyticity 
or unitarity properties of $\mathcal{A}(s,t)$, which remains the fundamental Lorentz-invariant 
object. The $b$-space formalism is thus an effective tool for analyzing 
the transverse structure of high-energy scattering without contradicting 
these fundamental principles.

Re-defining $\tau = \log s$, one can then show, after straightforward
algebra, that Eq.~(\ref{eq:pomeron-amplitude}) obeys a diffusion equation
with a linear term,
\begin{eqnarray}
   (\partial_{\tau}-\alpha'\nabla_b^2-\epsilon_{\mathcal{P}})\,\widetilde{T}(\tau,b)=0~,
   \label{eq:FKPP}
\end{eqnarray}
where the Laplacian in 2-dimensions is 
\begin{eqnarray}  \nabla_b^2=\partial_{bb}+\frac{1}{b}\partial_b~.
\end{eqnarray}
The diffusion equation draws an analogy with the Schrödinger equation in imaginary time, with $\tau$ playing the role of the logarithm of the scattering energy (the rapidity).

However,  the linear term $\epsilon_{\mathcal{P}}\widetilde{T}$ leads to power-law growth of amplitude, which ultimately violates both the Froissart bound \cite{Froissart:1961} and the unitarity at asymptotic energies. That is, the simple-pole amplitude grows faster than it spreads in $b$-space. In a sense, at high energies the simple-pole Pomeron-like amplitude Eq.(\ref{eq:pomeron-amplitude0}) is not appropriate to describe elastic scattering and must be replaced by something else. 
This behavior indicates that the linear evolution, although analytically tractable and phenomenologically useful at moderate energies, becomes insufficient to describe the dynamics at high energies where unitarity constraints become significant. To account for these effects and ensure a more realistic asymptotic behavior, it is natural to seek a nonlinear extension of Eq.~(\ref{eq:FKPP}) that can tame the unbounded growth of the amplitude. Motivated by the fact that at high energies the density of Pomeron increases and eventually reaches saturation, which in elastic scattering case can be interpreted as the unitarity constraint, we propose the following nonlinear generalization:
\begin{eqnarray}
   \left[\partial_{\tau}-\alpha'\nabla_b^2-\epsilon_{\mathcal{P}}\left(1+i\,\frac{\lambda}{\epsilon_{\mathcal{P}}}\,\widetilde{T}\right)\right]\,\widetilde{T}=0~,
   \label{Eq:EvolutionEquation_KKK}
\end{eqnarray}
and look for solutions of this equation.
Separating it into the real and imaginary parts we arrive at a set of two coupled partial differential equations
\begin{eqnarray}
\partial_{\tau}\widetilde{T}_R=\alpha'\nabla_b^2\,\widetilde{T}_R+\epsilon_{\mathcal{P}}\left(1-2\frac{\lambda}{\epsilon_{\mathcal{P}}}\,\widetilde{T}_I\right)\,\widetilde{T}_R~,
   \label{Eq:Real}
\end{eqnarray}
\begin{eqnarray}
\partial_{\tau}\widetilde{T}_I=\alpha'\nabla_b^2\,\widetilde{T}_I+\epsilon_{\mathcal{P}}\left(1-\frac{\lambda}{\epsilon_{\mathcal{P}}}\,\widetilde{T}_I\right)\,\widetilde{T}_I+\lambda\,\widetilde{T}_R^2~.
   \label{Eq:Imaginary}
\end{eqnarray}
We presented the above complex evolution equation for scattering amplitudes in a recent work \cite{Kakkad:2022ama}, based on Regge field theory, changing the standard approach in elastic scattering, turning it into an initial value problem. 

The imaginary part Eq.(\ref{Eq:Imaginary}) resembles the well-known Fisher-Komolgorov-Petrovsky-Piscunov equation  (FKPP) \cite{FKPP:1937_A,FKPP:1937_B}  with an additional nonlinear term from the real part which usually is subdominant. Disregarding the quadratic term $\widetilde{T}_R^2$ the imaginary part of the equation has some well-established properties, and under specific conditions it can be solved analytically \cite{Zeppetella:1979}.
This kind of equation was vastly explored in different scientific domains, biology, chemistry and physics. Recently, this equation has attracted interest in the high-energy physics community, since it maps the energy evolution described by an integro-differential equation for the gluon density \cite{Munier-Peschanski:2003}.  The imaginary part resembles the Balitsky-Kovchegov (BK) equation  \cite{Balitsky:1995ub,Kovchegov:1999yj,Marquet:2005qu} and the real part the Kovchegov-Szymanowski-Wallon (KSW) equation \cite{Odderon:2003}.

There is no justification for neglecting the quadratic term $\widetilde{T}_R^2$, which leads to a system of coupled partial differential equations. The initial conditions required to solve this system do not yield closed-form analytic solutions. In fact, analytic solutions are only possible under the assumption of a traveling wave with a specific speed, as shown in Ref.\cite{Zeppetella:1979}. In our previous work\cite{Kakkad:2022ama}, we began our analysis by using models defined in $b$-space as initial conditions for the evolution equation.

\subsection{Closed-form solution in the diffusionless limit}

 In the Regge framework, the diffusion term in $b$-space associated with a non-zero Pomeron slope $\alpha'$, first discussed by Gribov \cite{Gribov-diffusion}, is responsible for the transverse spreading of the interaction region and for the shrinkage of the diffraction cone. However, at sufficiently high energies, the scattering amplitude in impact parameter space is expected to approach the unitarity limit in the central region, small $b$, where nonlinear effects become dominant.

In this regime, the evolution is driven primarily by the interplay between the growth term $\epsilon_{\mathcal{P}}$ and the nonlinear saturation term $\lambda$, which locally unitarizes the amplitude. As a consequence, the role of transverse diffusion may become subleading in shaping the core of the interaction region, while still may contributing to its periphery.

The $\alpha' = 0$ limit considered in the present work should therefore be understood as an effective approximation that isolates this nonlinear-dominated regime. This simplification transforms the evolution equation into a local nonlinear equation in rapidity, allowing for an analytic solution in closed form. In this way, the essential features of saturation dynamics can be investigated without the additional complexity introduced by diffusion.

This interpretation is further supported by the fact that key qualitative features of the evolution, such as the emergence of a propagating front, can still be reproduced in the absence of an explicit diffusion term, as illustrated by the comparison with FKPP-type dynamics discussed below. A more complete treatment including both diffusion and nonlinear effects is left for future work.

In this diffusionless limit, the evolution equation reduces to a first-order nonlinear differential equation

\begin{eqnarray}
   \left[\frac{\partial}{\partial\tau}-\epsilon_{\mathcal{P}}\left(1+i\frac{\lambda}{\epsilon_{\mathcal{P}}}\,\widetilde{T}\right)\right]\,\widetilde{T}=0~
   \label{eq:Complex_Logistic}
\end{eqnarray}
with a complex analytic solution

\begin{eqnarray}
   \widetilde{T}(\tau,b)=\frac{\widetilde{T}_0(\tau_0,b)}{\Big(1+i\frac{\lambda}{\epsilon_{\mathcal{P}}}\widetilde{T}_0(\tau_0,b)\Big)\,e^{-\epsilon_{\mathcal{P}}(\tau-\tau_0)}-i\frac{\lambda\, }{\epsilon_{\mathcal{P}}}\widetilde{T}_0(\tau_0,b) }~,
   \label{eq:Analytic_Solution}
\end{eqnarray}
where $\widetilde{T}_0(\tau_0,b)$ is the $b$-profile at a given energy and $\tau_0$ playing the role of  initial value.  As mentioned in the introduction, the use of phenomenological profiles for  $\widetilde{T}(\tau_0, b)$  at a fixed energy is not a limitation, but a way to test the universality of our evolution equation. Once the initial profile is set, the energy dependence is fully determined by the nonlinear dynamics, much like the role of parton distribution functions in QCD, where nonperturbative input evolves through universal equations.

Although the evolution equation studied in this work is formulated directly at the level of the elastic amplitude in impact-parameter space, it is not introduced as a purely phenomenological ansatz. As shown in appendix B, it can be derived within an effective Reggeon Field Theory framework, where it emerges as the mean-field reduction of a nonlinear Reggeon hierarchy with triple-Pomeron interactions. The present approach should therefore be understood as an effective high-energy collective description, consistent with S-matrix unitarity and analyticity constraints, rather than as a fundamental QCD derivation.

In appendix B we show how to obtain Eq.(\ref{eq:Complex_Logistic}) from RFT, where the Pomeron intercept $\epsilon_{\mathcal{P}}$ is interpreted as an effective mass and the nonlinear term $\lambda$ is interpreted as the interaction coupling of the triple-Pomeron vertex. The formalism can be generalized to $\alpha' \neq 0$, although the analytic solution 
is then lost.

To find the solution Eq.(\ref{eq:Analytic_Solution}) we  change the function $\widetilde{T}(\tau,b)$ to $u(\tau, b)=1/\widetilde{T}(\tau,b)$ and the differential equation  transforms into a linear relaxation equation
\begin{eqnarray}
   \Big[\frac{\partial}{\partial\tau}+\epsilon_{\mathcal{P}}\Big]\,u(\tau, b)=-i\,\lambda~,
   \label{eq:Complex_Logistic_u}
\end{eqnarray}
with the  solution given by the sum of the homogeneous and the particular parts
\begin{eqnarray}
   u(\tau, b)=A(b)\,e^{-\epsilon_{\mathcal{P}}\,\tau}-i\,\frac{\lambda}{\epsilon_{\mathcal{P}}}~
   \label{eq:solution_u}
\end{eqnarray}
where $A(b)$ is an integration constant with respect to $\tau$ but with a $b$-dependence.
The amplitude is then 
\begin{eqnarray}
    \widetilde{T}(\tau, b)=\frac{1}{A(b)\,e^{-\epsilon_{\mathcal{P}\,}\tau}-i\,\frac{\lambda}{\epsilon_{\mathcal{P}}}}~. 
\end{eqnarray}
Using $\widetilde{T}(\tau_0,b)=\widetilde{T}_0(b)$ as initial condition we obtain $A(b)$, and as a consequence the solution Eq.(\ref{eq:Analytic_Solution}).

The logistic-type nonlinear evolution of the elastic amplitude in $b$-space can be mapped onto a dissipative (damped) system, where the nonlinear term functions as a regulator of the dynamics, ensuring unitarity, much like damping enforces equilibrium in mechanical systems. 
This invites an analogy with damped classical systems, such as RC circuits or overdamped harmonic oscillators, where damping plays the role of a mechanism that conducts the system toward equilibrium. In this dual representation, unitarization emerges either as nonlinear damping in $\widetilde{T}$ or as linear relaxation in $u$, providing both conceptual and computational insights into the dynamics of high-energy elastic scattering. This analogy may also open the door to analyzing stability, fixed points, and relaxation times using tools from dynamical systems theory.

Equation (\ref{eq:Analytic_Solution}) is remarkably simple and can quantitatively describe the experimental data from the ISR to the LHC energies (also the cosmic ray energies) using only two physical quantities. However, the values of the parameter $\rho$, which is represented by  the  ratio between the real and imaginary  parts at $|t|=0$,   as coming from the experimental analysis are not sufficiently described and a reinterpretation of this quantity may be necessary. To account for a flexibility in the description of the real part of the amplitude we can simply promote $\epsilon_\mathcal{P}$ to a complex effective mass $\tilde{\epsilon}=\epsilon_R+i\epsilon_I$
\begin{eqnarray}
   \left[\frac{\partial}{\partial\tau}-\widetilde{\epsilon}\,\left(1+i\,\frac{\lambda}{\widetilde{\epsilon}}\,\widetilde{T}\right)\right]\,\widetilde{T}=0~,
   \label{eq:Complex_Logistic_2}
\end{eqnarray}
introducing one more parameter in the equation, holding the same analytic structure for the complex solution. It is interesting to express the real and imaginary parts of the differential equation to understand their behavior
\begin{eqnarray}
\frac{\partial\,\widetilde{T}_R}{\partial\tau}=\epsilon_{R}\,\left(1-\frac{2\,\lambda}{\epsilon_{R}}\,\widetilde{T}_I\right)\,\widetilde{T}_R+\epsilon_{I}\,\widetilde{T}_I
\label{Eq:RealLogistic}
\end{eqnarray}
\begin{eqnarray}
\frac{\partial\,\widetilde{T}_I}{\partial\tau}=\epsilon_{R}\,\left(1-\frac{\lambda}{\epsilon_{R}}\,\widetilde{T}_I\right)\,\widetilde{T}_I+\epsilon_{I}\,\left(1-\frac{\lambda}{\epsilon_{I}}\,\widetilde{T}_R\right)\,\widetilde{T}_R~.
   \label{Eq:ImaginaryLogistic}
\end{eqnarray}
Taking $\epsilon_{I}\to 0$  and  $\lambda\,\widetilde{T}_R^2\to 0$,  Eqs. (\ref{Eq:RealLogistic}) and (\ref{Eq:ImaginaryLogistic})  approximates to
\begin{eqnarray}
\frac{\partial\,\widetilde{T}_R}{\partial\tau}\simeq\epsilon_{R}\,\left(1-\frac{2\,\lambda}{\epsilon_{R}}\,\widetilde{T}_I\right)\,\widetilde{T}_R
   \label{Eq:RealLogistic_Approx}
\end{eqnarray}
\begin{eqnarray}
\frac{\partial\,\widetilde{T}_I}{\partial\tau}\simeq\epsilon_{R}\,\left(1-\frac{\lambda}{\epsilon_{R}}\,\widetilde{T}_I\right)\,\widetilde{T}_I~,
   \label{Eq:ImaginaryLogistic_approx}
\end{eqnarray}
respectively.

The evolution of the imaginary part of the amplitude in impact parameter space, $\widetilde{T}_I$, follows a logistic growth pattern for a given initial profile. This behavior naturally emerges from the nonlinear term and reflects a saturation mechanism as energy increases. At low energies, the amplitude grows linearly with $\tau$, passing by a region where nonlinear effects become significant, leading to a saturation of the amplitude.

\begin{figure}[H]
    \centering
    \includegraphics[width=0.95\linewidth]{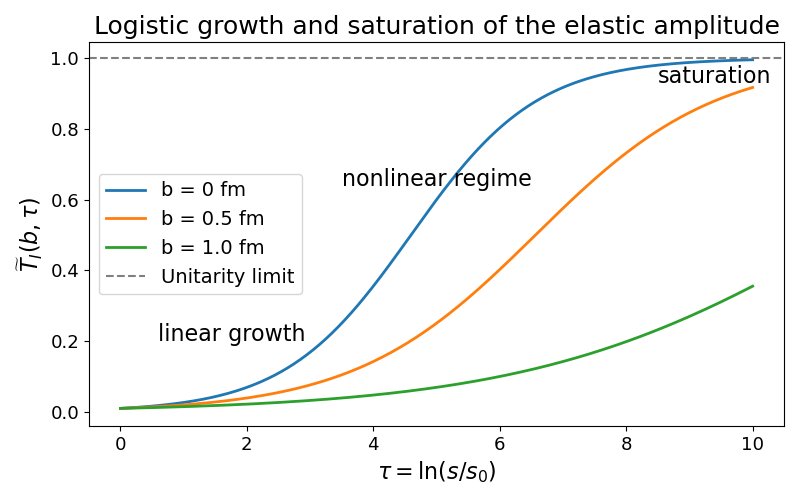}
    \caption{Evolution of the pure logistic equation for three different initial conditions corresponding to the imaginary part of the elastic amplitude 
$\widetilde{T}(\tau,b)$ with three different values of $b$ as function of $\tau$. The amplitude grows linearly at low energies, but saturates due to nonlinear effects, in analogy with gluon saturation in QCD. The plateau corresponds to the unitarity bound.}
    \label{fig:b-profile}
\end{figure}

The saturation effect can be interpreted in the language of Regge theory by associating $\widetilde{T}_I$ with the Pomeron density profile in $b$-space. As the energy increases, the Pomeron density also increases, until nonlinear interactions among Pomerons become important and lead to a limiting behavior, effectively ''absorbing" Pomerons and saturating the density. In Fig.~\ref{fig:b-profile}, we display the evolution of the pure logistic equation, which corresponds to the imaginary part of the amplitude 
$\widetilde{T}(\tau,b)$. For any fixed value of the impact parameter 
$b$, the amplitude undergoes an initial linear growth at low 
$\tau$, followed by a nonlinear transition region, and asymptotically approaches a saturation plateau dictated by the unitarity constraint.

Such a mechanism resembles the gluon saturation phenomenon in QCD, as described by the nonlinear evolution equations such as BK \cite{Balitsky:1995ub,Kovchegov:1999yj,Marquet:2005qu} and B-JIMWLK \cite{Balitsky:1995ub,JalilianMarian:1997jx,JalilianMarian:1997gr,JalilianMarian:1997dw,Kovner:2000pt,Kovner:1999bj,Weigert:2000gi,Iancu:2000hn,Ferreiro:2001qy}, where gluon recombination slows down the growth of the gluon distribution at large rapidity, when gluons carry small  momentum fraction of the incident hadron (small-$x$ physics). This connection is not entirely unexpected, as gluon saturation arises from ladder-type gluon interactions, precisely the building blocks of the QCD Pomeron \cite{Forshaw:1997dc}.  In fact, it has been shown that the JIMWLK evolution can be reinterpreted as a Reggeon field theory in its own right, where Pomerons and their interactions emerge dynamically from QCD \cite{Kovner:2005qj,Altinoluk:2009je}.

An interesting aspect of Eq.(\ref{eq:Complex_Logistic}) is the stability of the traveling wave due to the interplay between the nonlinear and the dispersive linear terms, behaving as a kink soliton solution. The two  solutions are: ($\widetilde{T}=0$ unstable) and ($\widetilde{T}=i\,\frac{\epsilon_\mathcal{P}}{\lambda}$ stable). Assuming a non zero initial condition $T_0$, we can describe a kink soliton obtained from the solution Eq.(\ref{eq:Analytic_Solution}) passing from $\tau=-\infty$ to $\tau\to \infty$. At asymptotic energies the amplitude is purely imaginary and the scattering is dominated by the inelastic processes. On the other hand, if we take the complex parameter $\tilde{\epsilon}$ in Eq.(\ref{eq:Analytic_Solution}) the asymptotic solution becomes 
\begin{eqnarray}
\lim_{\tau\to\infty}\widetilde{T}(\tau,b)\to\frac{-\epsilon_{I}+i\,\epsilon_R}{\lambda}
\label{Eq:Solution_Asymptotic}
\end{eqnarray}
 and the real part of the elastic processes is still asymptotically preserved.

Note that the diffusion term is the main difference between the QCD evolution equations and the evolution equation (\ref{eq:Complex_Logistic_2}). 
In perturbative QCD, for instance, the BK equation, mapped as FKPP equation, the diffusive term is usually attributed to the propagating front of the traveling wave. However, FKPP equation also presents the non-linear term.

\begin{figure}[H]
    \centering
    \includegraphics[width=0.95\linewidth]{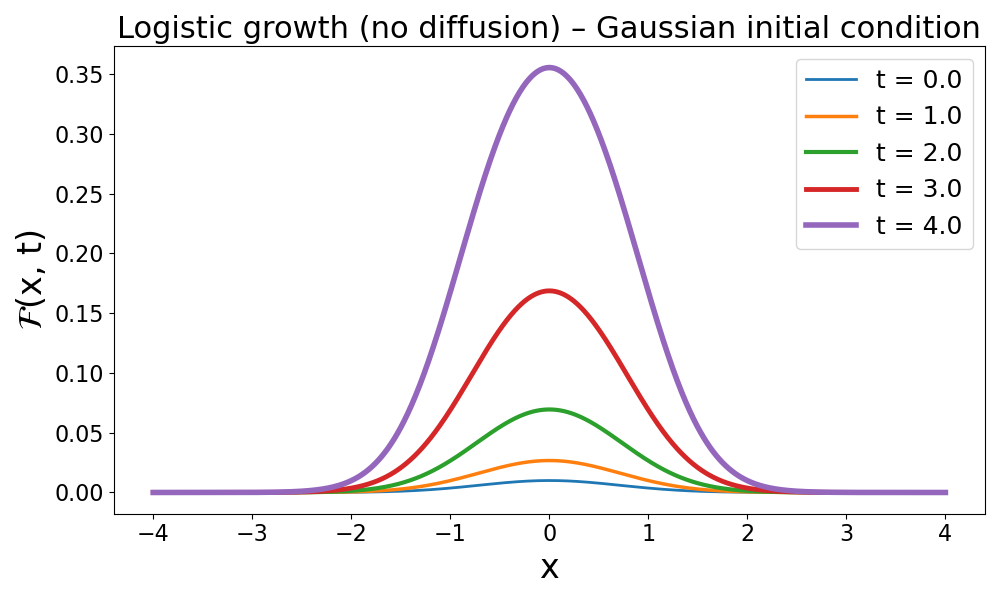}
    \includegraphics[width=0.95\linewidth]{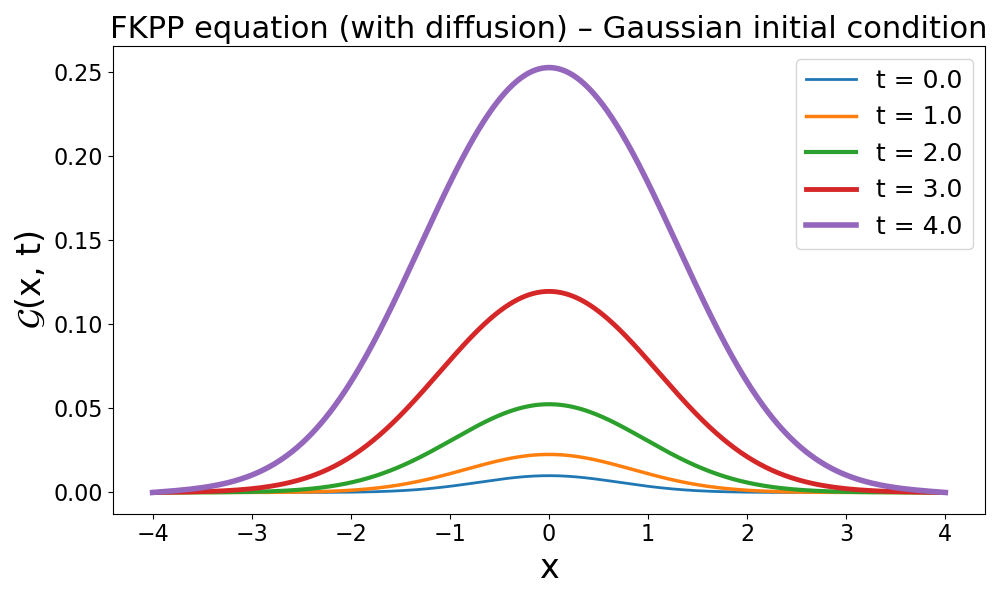}

    \caption{We compare pure logistic and FKPP solutions using a pure gaussian as initial condition. The upper panel shows the simple logistic equation without the diffusion term, whereas the lower shows the FKPP equation with the diffusion coefficient set to  0.1. Notice that the diffusive part spreads out the solution faster than the diffusionless part. Increasing the diffusion parameter increases the speed of the spread.  }
    \label{fig:diffusion_nodiffusion}
\end{figure}

To understand similarities and differences we compare a real logistic solution with FKPP type using a pure Gaussian as initial condition in both cases. For  FKPP we set the diffusion coefficient to $\mathcal{D}=0.1$. Of course, increasing this coefficient we increase the speed of the wave front. We show, that even without a diffusive term in logistic equation we still have a propagating wave front, due to the tail of the Gaussian provided as initial condition, with the advantage of having an analytic solution in closed form. The equations are:
\begin{eqnarray}
\frac{\partial \mathcal{F}(x,t)}{\partial t}=\epsilon\,\left(1-\frac{\lambda}{\epsilon}\,\mathcal{F}(x,t)\right)\,\mathcal{F}(x,t)  \end{eqnarray}
logistic equation with an analytic solution
\begin{eqnarray}
    \mathcal{F}(x,t)=\frac{\mathcal{F}_0(x)\,e^{\epsilon\,t}}{1+\frac{\lambda}{\epsilon}\,\mathcal{F}_0(x)\Big(e^{\epsilon\,t}-1\Big)}  ~,
\end{eqnarray}
and the FKPP equation
\begin{eqnarray}
\frac{\partial \mathcal{G}(x,t)}{\partial t}=\mathcal{D}\,\frac{\partial^2 \mathcal{G}(x,t)}{\partial x^2}+\epsilon\,\left(1-\frac{\lambda}{\epsilon}\,\mathcal{G}(x,t)\right)\,\mathcal{G}(x,t)  ~
\end{eqnarray}
solved numerically. The initial conditions  are respectively:
\begin{eqnarray}\mathcal{F}_0(x)=\mathcal{F}(x,t_0)=A\,e^{-\frac{x^2}{\beta}}~,
\end{eqnarray}
and
\begin{eqnarray}
~~\mathcal{G}_0(x)=\mathcal{G}(x,t_0)=A\,e^{-\frac{x^2}{\beta}}~.
\end{eqnarray}
 The common parameters are : $\epsilon=1$, $\lambda=1$, $A=0.01$, $\beta=1$ and ($\mathcal{D}=0.1$ exclusively for FKPP). In fig. \ref{fig:diffusion_nodiffusion} we compare the two equations using similar conditions. We plot the time  evolution from 0 to 4 in equally spaced steps, where we see qualitatively the similar behavior.

\clearpage

\section{Mathematical properties and asymptotic theorems}

In this section, we demonstrate that the proposed evolution equation satisfies the fundamental mathematical and asymptotic properties expected of high-energy scattering amplitudes. We show explicitly that the equation preserves unitarity, rigourosly respects analyticity and the associated dispersion relations, and admits a unique, well-defined solution under physically motivated conditions. These rigorous results establish that the framework is not only phenomenologically motivated but also mathematically consistent with the key theorems that govern the high-energy regime of quantum field theory.

\subsection{Unitarity}

Throughout this work, the physical elastic S-matrix in impact-parameter space
is defined as
\begin{eqnarray}
S(s,b) = 1 + i \widetilde{T}^{(p)}(s,b)~    ,
\end{eqnarray}
the physical amplitude is denoted as $\widetilde{T}^{(p)}(s,b)$ and all unitarity constraints refer to this definition.
With this convention, elastic unitarity implies 
\begin{equation}
|S(s,b)|^2 \le 1 ,
\label{Eq:Unitarity_S_matrix}
\end{equation}
which defines the unit disk in the
$\big (\widetilde{T}^{(p)}_{I}(s,b)$ --- $\widetilde{T}^{(p)}_{R}(s,b)\big)$ complex plane.

Equivalently, elastic unitarity at fixed impact parameter can be written in the standard S-matrix form
\begin{equation}
2\,\mathrm{Im}\,\widetilde{T}^{(p)}(s,b)
= |\widetilde{T}^{(p)}(s,b)|^2 + G_{\text{inel}}(s,b),
\end{equation}
where $G_{\text{inel}}(s,b)\ge 0$ is the inelastic overlap function.
The inequality $|S(s,b)| \le 1$ is therefore fully equivalent to the usual unitarity condition.

The boundary of this disk is given by
\begin{equation}
\big[\widetilde{T}^{(p)}_{I}(s,b)-1\big]^2
+\big[\widetilde{T}^{(p)}_{R}(s,b)\big]^2 = 1 .
\label{Eq:S_matrix_boundary}
\end{equation}
This equation represents a circle centered at $(0,1)$ with unit radius.

Solving Eq.~(\ref{Eq:S_matrix_boundary}) for the imaginary part of the amplitude,
one finds
\begin{eqnarray}
\widetilde{T}^{(p)}_{I\,\pm}(s,b)=1\pm\sqrt{1-\big[\widetilde{T}^{(p)}_{R}(s,b)\big]^2}
\label{Eq:S_matrix_equality} ~.
\end{eqnarray}

The upper branch corresponds to
\begin{equation}
1 \le \widetilde{T}^{(p)}_{I+}(s,b) \le 2 ,
\end{equation}
while the lower branch satisfies
\begin{equation}
0 \le \widetilde{T}^{(p)}_{I-}(s,b) \le 1 .
\label{Ineq:Unitarity_phys}
\end{equation}
The latter is the physically relevant branch, which ensures
$|S(s,b)| \le 1$ at each impact parameter.

\vspace{0.2cm}
To analyze the nonlinear energy evolution, we introduce an auxiliary amplitude
$\widetilde{T}(s,b)$ governed by the complex logistic equation
(\ref{eq:Complex_Logistic}).
The auxiliary amplitude has no direct unitarity interpretation.
The physical amplitude is obtained through the constant rescaling
\begin{equation}
\widetilde{T}^{(p)}(s,b)
= \frac{\lambda}{\epsilon_{\mathcal P}}\,\widetilde{T}(s,b) ,
\label{Eq:T_rescaling}
\end{equation}
which leaves the structure of the evolution equation unchanged.

The evolution equation (\ref{eq:Complex_Logistic}) admits two fixed points for
the auxiliary amplitude,
$\widetilde{T}_I^{\rm (min)}=0$ and
$\widetilde{T}_I^{\rm (max)}=\epsilon_{\mathcal P}/\lambda$.
Using Eq.~(\ref{Eq:T_rescaling}), the corresponding fixed points of the
physical amplitude are
\begin{equation}
\widetilde{T}^{(p)}_I = 0
\quad \text{and} \quad
\widetilde{T}^{(p)}_I = 1 ,
\end{equation}
so that the nonlinear evolution dynamically drives the amplitude toward the
unitarity bound~(\ref{Ineq:Unitarity_phys}) at fixed impact parameter.

\vspace{0.2cm}

The exact solution of the evolution equation (\ref{eq:Analytic_Solution}) for the physical amplitude can
be written as
\begin{equation}
\widetilde{T}^{(p)}(\tau,b)
= \frac{\widetilde{T}_0(b)}
{\left(1+i\,\widetilde{T}_0(b)\right)e^{-\epsilon_{\mathcal P}(\tau-\tau_0)}
- i\,\widetilde{T}_0(b)} ,
\end{equation}
where $\widetilde{T}_0(b)$ is an arbitrary complex initial profile.

The corresponding S-matrix takes the exact form
\begin{equation}
S(\tau,b)
= \frac{S_0(b)\,e^{-\epsilon_{\mathcal P}(\tau-\tau_0)}}
{S_0(b)\left(e^{-\epsilon_{\mathcal P}(\tau-\tau_0)} - 1\right)+1},
\label{Eq:S_Mobius}
\end{equation}
with
\begin{equation}
S_0(b) = 1 + i\,\widetilde{T}_0(b).
\end{equation}

Equation~(\ref{Eq:S_Mobius}) defines a fractional linear (Möbius)
transformation acting on the initial S-matrix. We now assume
\begin{equation}
\Re(\epsilon_{\mathcal P}) > 0 ,
\label{Eq:epsilon_condition_appendix}
\end{equation}
so that
\begin{equation}
\left|e^{-\epsilon_{\mathcal P}(\tau-\tau_0)}\right|
= e^{-\Re(\epsilon_{\mathcal P})(\tau-\tau_0)} \le 1 .
\end{equation}

For fixed $b$, the evolution thus maps $S_0(b)$ into $S(\tau,b)$ through a
Möbius transformation with contraction factor of modulus smaller than unity.
It follows that the unit disk in the complex $S$-plane is mapped into itself.
Therefore, if the initial condition is physical, i.e.
\begin{equation}
|S_0(b)| \le 1 ,
\end{equation}
then the solution satisfies
\begin{equation}
|S(\tau,b)| \le 1
\qquad \text{for all } \tau \text{ and } b .
\end{equation}

Hence, the exact solution of the nonlinear evolution equation preserves
unitarity for arbitrary complex initial profiles, provided that
$\Re(\epsilon_{\mathcal P})>0$.

\subsection{ Froissart-Martin limit}

Another key ingredient in hadronic scattering is the short-range behavior of  strong interactions. Together with unitarity, these two ingredients allow to derive the Froissart theorem, establishing an upper bound for the asymptotic growth of the total cross section \cite{Froissart:1961} leading to the bound
\begin{eqnarray}
\sigma_{\rm tot.}(\tau(s)\to \infty)\leq \mathcal{C}\,\tau^2=\mathcal{C}\,\log^2 s ~.
    \label{Eq:Froissart}
\end{eqnarray}
From local and massive field theory Martin \cite{Martin:346321} proved that  constant  $\mathcal{C}=\frac{\pi}{m_{\pi}^2}\simeq 172 $ mb where $m_{\pi}$ is the pion mass. 

From a theoretical perspective, 
the total cross section is determined by the extrapolation of the imaginary amplitude at the optical point. Up to now, Regge phenomenology was a successful approach to describe the energy dependence of the total elastic scattering of hadrons. In the seminal work from Donnachie- Landshoff, the Regge trajectories and the Pomeron trajectory were able to describe the hadronic scattering data in a broad energy range \cite{Donnachie:1992ny}. However, the positive power in the energy dependence due to the single-Pomeron intercept violates the Froissart bound. The unitarization of such a model was captured later by the introduction of a double Pomeron interaction.  

We propose an evolution equation that, given a profile amplitude in $b$ at a fixed energy, is able to describe all  energy dependence of the scattering processes up to one physical quantity to be determined. To assess whether this evolution respects the  Froissart-Martin bound we use the optical theorem and Fourier transform the imaginary part of the scattering amplitude from the $b$ to the $t$ space and we take the limit where $t\to 0$. 
\begin{eqnarray}
    \sigma_{\rm tot.}(\tau)=2\,T_I(\tau,t=0)=4\pi\,\int_0^{\infty} b\,db\,J_0(0)\,\widetilde{T}_{I}(\tau,b) ~.
    \label{Eq:Total_Cross}
\end{eqnarray}
For specific forms of function $\widetilde{T}_I(\tau,b)$ the equation above can be calculated in closed form. Geometrically, the interaction range is described by the imaginary profile in $b$-space given by the initial conditions, provided by models at fixed energy. In many models, the imaginary part of the scattering amplitude is one order of magnitude greater than the real part. For this reason, many authors simplify their models  focusing only on the aspects of the imaginary amplitude. In general, the behavior of the profiles is smooth and decrease with increasing $b$. In what follows, to obtain an analytic solution for the total cross section, we consider two extreme purely imaginary profiles: the Gaussian profile and the exponential one. The Gaussian profile is given by
\begin{eqnarray}
   T_0(\tau_0,b)=i\,B\,e^{-\beta\,b^2}~,
   \label{eq:Profile_Gaussian}
\end{eqnarray}
with $B$ and  $\beta$ given as constant parameters.  The amplitude in $t$-space has an analytic solution and the total cross sections is given  (the integral is calculated in appendix \ref{Appendix_D})
\begin{eqnarray}
   \sigma_{\rm tot.}^{\rm Gauss}(\tau)&=& 4\,\pi\,(\hbar\,c)^2\int_0^{\infty} b\,db\,\frac{B\,e^{-\beta\,b^2}}{\left(1-\frac{\lambda}{\epsilon_{\mathcal{P}}}B\,e^{-\beta\,b^2}\right)\,e^{-\epsilon_{\mathcal{P}}(\tau-\tau_0)}+\frac{\lambda\, }{\epsilon_{\mathcal{P}}}B\,e^{-\beta\,b^2} }~, \nonumber \\
   &=&4\,\pi\,(\hbar\,c)^2\, \frac{e^{\epsilon_{\mathcal{P}}(\tau-\tau_0)}\,\log\left(\frac{\epsilon_{\mathcal{P}}-B\,\lambda+B\,\lambda\,e^{\epsilon_{\mathcal{P}}(\tau-\tau_0)}}{\epsilon_{\mathcal{P}}} \right)}{2\,\beta\,\lambda\,(e^{\epsilon_{\mathcal{P}}(\tau-\tau_0)}-1)} ~.
   \label{eq:Analytic_Solution_Gaussian}
\end{eqnarray}
For an exponential profile function 
\begin{eqnarray}
   T_0(\tau_0,b)=i\,B'\,e^{-\beta'\,b}~,
   \label{eq:Profile_Exponential}
\end{eqnarray}
with $B'$ and $\beta'$ as constant parameters
the total cross section is
\begin{eqnarray}
   \sigma_{\rm tot.}^{\rm exp.}(\tau)&=& 4\,\pi\,(\hbar\,c)^2\, \frac{\epsilon_{{\mathcal{P}}}\,e^{\epsilon_{\mathcal{P}}(\tau-\tau_0)}\,Li_2\Big(\frac{-B'(e^{\epsilon_{\mathcal{P}}(\tau-\tau_0)}-1)\lambda}{\epsilon_{\mathcal{P}}} \Big)}{\beta'^2\,\lambda\,(e^{\epsilon_{\mathcal{P}}(\tau-\tau_0)}-1)} ~.
\label{eq:Analytic_Solution_Exponential}
\end{eqnarray}

For the Gaussian profile,  the asymptotic limit when $\tau \to \infty$, is given by
\begin{eqnarray}
 \lim_{\tau\to \infty}\frac{\sigma_{\rm tot.}^{\rm Gauss}(\tau)}{\tau}= \frac{\epsilon_{\mathcal{P}}^2 }{2\,\beta\,\lambda}~,
    \label{Eq:ratio_Gauss}
\end{eqnarray}
showing that the total cross section will behave as simply as $\tau=\log s$ at asymptotic energies, whereas for the exponential profile we have
\begin{eqnarray}
 \lim_{\tau\to \infty}\frac{\sigma_{\rm tot.}^{\rm exp.}(\tau)}{\tau^2}= \frac{\epsilon_{\mathcal{P}}^3 }{2\,\beta^2\,\lambda}~,
    \label{Eq:ratio_Exp}
\end{eqnarray}
which means that the total cross section behaves as ($\tau^2=\log^2 s$). In the second case, the coefficient $\frac{\epsilon_{\mathcal{P}}^3 }{2\,\beta^2\,\lambda} << \mathcal{C}$ for all the reasonable values of the parameters $\epsilon_{\mathcal{P}}$ , $\beta$ and $\lambda$, meaning that it satisfies Martin's bound coefficient.  

This simple exercise shows that  profiles  with functional forms from Gaussian to exponential $b$, lead to the total cross sections consistent with the Froissart-Martin limit, since Gaussian profiles fall much faster than exponential profiles with increasing $b$. As a conclusion, functions with $b$-dependence larger than exponential could lead to a violation of the Froissart bound, being considered long-range interactions.

\subsubsection{Regge interpretation and Froissart saturation}

The present framework does not assume a linear Regge trajectory of the form
$\alpha(t)=\alpha(0)+\alpha' t$, and in particular we set $\alpha'=0$.
In standard Regge phenomenology, the energy dependence of the total cross
section is governed by the intercept $\alpha(0)$ of the leading trajectory,
while the slope $\alpha'$ controls the transverse structure of the
interaction, such as the shrinkage of the forward diffraction peak and the
logarithmic growth of the interaction radius. In the present framework, the
transverse growth of the interaction region does not originate from Regge
diffusion associated with a nonzero $\alpha'$, but emerges dynamically from
the nonlinear energy evolution of the scattering amplitude in
impact-parameter space. The parameter $\epsilon_{\mathcal P}$ controls the
effective pre-unitarization energy growth, while the nonlinear dynamics
enforces saturation at fixed impact parameter.

In the language of complex angular momentum, saturation of the Froissart
bound is known to correspond to an enhanced singularity at $\ell=1$ in the
$t=0$ channel. It has long been established in the unitarized Regge
literature that eikonal or similar unitarization mechanisms can transform
a leading Regge singularity into an effective higher-order pole at
$\ell=1$, producing $\log^2 s$ behavior for the total cross section.
Although a full analytic continuation of the present amplitude to complex
$\ell$ is beyond the scope of this work, the asymptotic behavior derived
here is fully consistent with this Regge interpretation of Froissart
saturation and with analyticity in spin.

\subsection{Forward real amplitude}

 The parameter $\rho$ is defined as the ratio between the real and imaginary
parts of the forward elastic scattering amplitude in $t$-space
\begin{equation}
\rho(s) \equiv
\frac{T_R(\tau,t=0)}{T_I(\tau,t=0)} .
\end{equation}
While the imaginary part at $t=0$ is directly related to the total cross
section through the optical theorem, the determination of the real part
requires additional dynamical input.

In order to estimate the possible range of $\rho$ within the present
framework, we adopt a simplified model for the initial complex profile
$\widetilde{T}_0(\tau_0,b)$, which allows for an analytic calculation of
the forward real amplitude. As discussed in
Refs.~\cite{Kohara:2017ats, Kohara:2018wng}, the real and imaginary parts of
the scattering amplitude generally exhibit different $t$-dependences.
However, since the present analysis is restricted to the forward region
$t\simeq 0$, we assume for simplicity that both components share the same
functional dependence in this limit.

As in the previous section, if the profile is a complex function with a Gaussian $b$-dependence 
\begin{eqnarray}
   T_0(\tau_0,b)=(A+i\,B)\,e^{-\beta\,b^2}~,
   \label{eq:Complex_Profile_Gaussian}
\end{eqnarray}
the Fourier transform of the real amplitude at $t=0$ is

\begin{eqnarray}
   &&T_{R}(\tau,t=0)= \int_0^{\infty} b\,db\,\,\widetilde{T}_{0R}(\tau,b)~ \\
   &&=\frac{\epsilon_{\mathcal{P}}\,e^{\epsilon_{\mathcal{P}}\,\Delta\tau}}{4\beta\lambda\,(e^{\epsilon_{\mathcal{P}}\,\Delta\tau}-1)}\left[\pi+2\tan^{-1}\left(\frac{\epsilon_{\mathcal{P}}-B\,\lambda+B\lambda\,e^{\epsilon_{\mathcal{P}}\Delta\tau}}{A\,\lambda(1-e^{\epsilon_{\mathcal{P}}\Delta\tau})}\right)\right]~. \nonumber
   \label{eq:Real_Analytic_Solution_Gaussian}
\end{eqnarray}
and the imaginary part is 
\begin{eqnarray}
   &&T_{I}(\tau,t=0)= \int_0^{\infty} b\,db\,\,\widetilde{T}_{0I}(\tau,b)~\\
   &&=\frac{\epsilon_{\mathcal{P}}\,e^{\epsilon_{\mathcal{P}}\,\Delta\tau}}{4\,\beta\,\lambda(e^{\epsilon_{\mathcal{P}}\Delta\tau}-1)}\log\left[1+\frac{2B\lambda}{\epsilon_{\mathcal{P}}}\Big(e^{\epsilon_{\mathcal{P}}\,\Delta\tau}-1\Big)+(A^2+B^2)\frac{\lambda^2}{\epsilon_{\mathcal{P}}^2}\Big(e^{\epsilon_{\mathcal{P}}\,\Delta\tau}-1\Big)^2\right]~. \nonumber
   \label{eq:Imaginary_Analytic_Solution_Gaussian}
\end{eqnarray}
The solution of the above integrals are give in appendix \ref{Appendix_D}.
The ratio of the real and imaginary parts at the origin becomes $\rho$
\begin{eqnarray}
\rho_{\rm gauss}(\tau)=\frac{\pi+2\tan^{-1}\Big(\frac{\epsilon_{\mathcal{P}}-B\,\lambda+B\lambda\,e^{\epsilon_{\mathcal{P}}\Delta\tau}}{A\,\lambda(1-e^{\epsilon_{\mathcal{P}}\Delta\tau})}\Big)}{\log\left[1+\frac{2B\lambda}{\epsilon_{\mathcal{P}}}\Big(e^{\epsilon_{\mathcal{P}}\,\Delta\tau}-1\Big)+(A^2+B^2)\frac{\lambda^2}{\epsilon_{\mathcal{P}}^2}\Big(e^{\epsilon_{\mathcal{P}}\,\Delta\tau}-1\Big)^2\right]}
    \label{Eq:rho_gaussian}
\end{eqnarray}
Taking the limit when $\tau$ goes to infinity 
\begin{eqnarray}
\lim_{\tau\to\infty}\rho_{\rm gauss}(\tau)\to 0 ~.
    \label{Eq:rho_gauss_infinity}
\end{eqnarray}
From Eq.(\ref{Eq:rho_gauss_infinity}) it becomes clear that the energy dependence of $\rho$ depends on the magnitude of the real and imaginary profiles as well as  the parameters $\lambda$ and $\epsilon_{\mathcal{P}}$.
More realistic and physically acceptable profiles may keep similar properties. 

\subsection{Uniqueness}

Equation (\ref{eq:Complex_Logistic}) provides a unique solution for each different initial condition. To prove the uniqueness it is sufficent to prove that it is Lipshitz continous, which is an essential condition for  Picard–Lindelöf theorem which guarantees the existence of uniqueness.
Let's recall Lipshitz theorem: 

{\it A function $f(z)$ is called Lipschitz continuous on a domain $\mathcal{D} \subset \mathbb{C}$ if there exists a constant $L>0$  such that for all $z_1$ and $z_2$ $\in~ \mathbb{C}$ we have
\begin{eqnarray}
    |f(z_1)-f(z_2)|\leq L|z_1-z_2|~.
\end{eqnarray}}

From Eq.(\ref{eq:Complex_Logistic_2}) we can rewrite the differential equation as
\begin{eqnarray}
   \frac{\partial\widetilde{T}}{\partial\tau}(\tau,b)=f(\widetilde{T}(\tau,b))~,
\label{eq:Complex_Logistic_Lipschitz}
\end{eqnarray}
where 
\begin{eqnarray}
f(\widetilde{T})=\epsilon_{\mathcal{P}}\widetilde{T}+i\lambda\widetilde{T}^2~,    
\end{eqnarray}
with $f(\widetilde{T})$ being our field of $\widetilde{T}\equiv\widetilde{T}(\tau,b)$.
\bigskip

{\it Theorem: The function $f(\widetilde{T})$ is Lipschitz continuous.}

Proof: Let's take two point $f(\widetilde{T}_1)$ and $f(\widetilde{T}_2)$ 

\begin{eqnarray}
 f(\widetilde{T}_1)-f(\widetilde{T}_2)&=&\epsilon_{\mathcal{P}}(\widetilde{T}_1-\widetilde{T}_2)+i\lambda\,(\widetilde{T}_1^2-\widetilde{T}_2^2) \nonumber \\
&&=\Big(\epsilon_{\mathcal{P}}+i\lambda\,(\widetilde{T}_1+\widetilde{T}_2)\Big)\,(\widetilde{T}_1-\widetilde{T}_2)
\end{eqnarray}
As a consequence the norm of the above equation is 
\begin{eqnarray}
 &&|f(\widetilde{T}_1)-f(\widetilde{T}_2)|=|\epsilon_{\mathcal{P}}+i\lambda\,(\widetilde{T}_1+\widetilde{T}_2)|\,|\widetilde{T}_1-\widetilde{T}_2|
 \end{eqnarray}
 and the triangular inequality gives
 \begin{eqnarray}
 |f(\widetilde{T}_1)-f(\widetilde{T}_2)|\leq \Big(|\epsilon_{\mathcal{P}}|+|\lambda|\,|\widetilde{T}_1+\widetilde{T}_2|\Big)\,|\widetilde{T}_1-\widetilde{T}_2|~.
\end{eqnarray}
Assuming $\widetilde{T}_1$ and $\widetilde{T}_2$ inside a bounded set $|\widetilde{T}_1|,|\widetilde{T}_2| \leq M$, we have $|\widetilde{T}_1+\widetilde{T}_2|\leq 2M$ that implies 
\begin{eqnarray}
 |f(\widetilde{T}_1)-f(\widetilde{T}_2)|\leq \Big(|\epsilon_{\mathcal{P}}|+2|\lambda|\,M\Big)\,|\widetilde{T}_1-\widetilde{T}_2|~,
\end{eqnarray}
where $L=|\epsilon_{\mathcal{P}}|+2|\lambda|\,M$. We thus prove that 
\begin{eqnarray}
    |f(\widetilde{T}_1)-f(\widetilde{T}_2)|\leq L\,|\widetilde{T}_1-\widetilde{T}_2|
\end{eqnarray}
and the function $f$ is Lipschitz continuous. 
Since the vector field $f(\widetilde{T}) = \epsilon_{\mathcal{P}}\widetilde{T} + i\lambda\widetilde{T}^2$ is continuous and locally Lipschitz continuous on $\mathbb{C}$, the Cauchy-Lipschitz (Picard–Lindelöf) theorem applies.  
Therefore, for any given initial condition $\widetilde{T}(\tau_0,b) = \widetilde{T}_0(b)$, there exists a unique solution $\widetilde{T}(\tau,b)$ to the differential equation in a neighborhood of $\tau_0$.  
In addition, because the vector field has polynomial growth and no singularities, the solution can be extended globally for all values of $\tau$.

\begin{figure}[H]
    \centering
    \includegraphics[width=0.8 \linewidth]{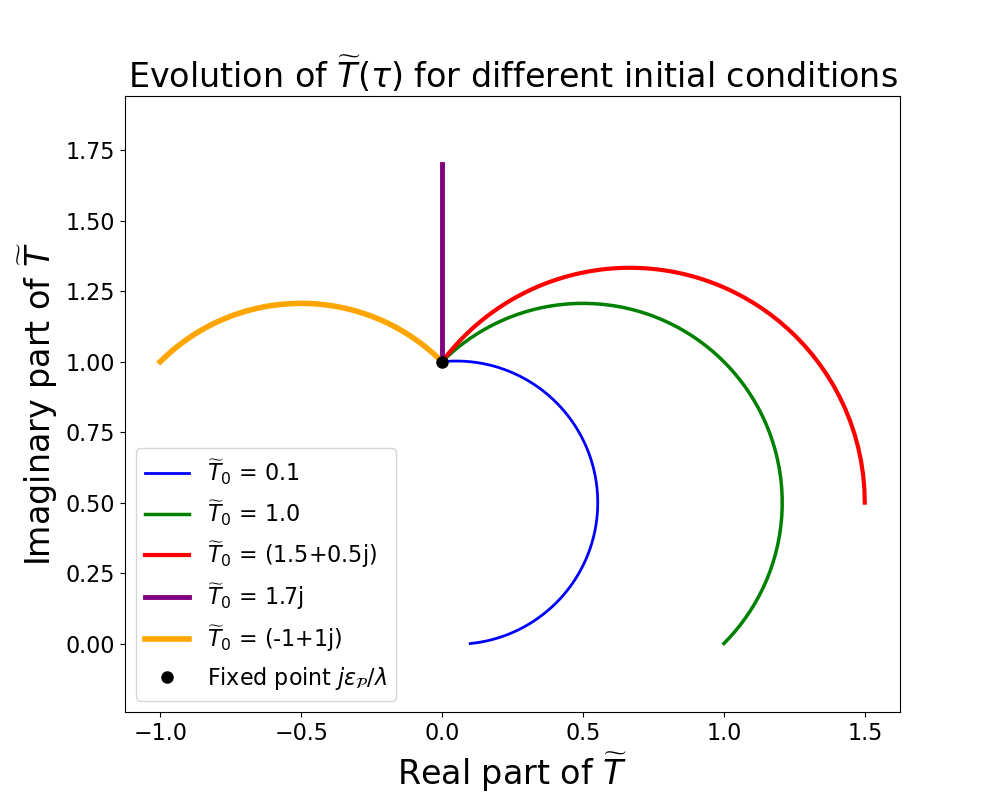}
    \caption{The figure shows several solutions $\widetilde{T}(\tau)$ corresponding to different complex initial conditions. The curves describes the $\tau$ evolution. 
    Despite the divergent initial trajectories as initial conditions in the complex plane, all the solutions converge asymptotically to the same value $i\epsilon_{\mathcal{P}}/\lambda$. We chose the parameters: $\epsilon_{\mathcal{P}}$ = 0.1, $\lambda=0.1$ and $\tau_0 =1
$.    }
    \label{fig:analytical_solutions}
\end{figure}

However, if we reach an infinity value of $\tau$ it is impossible to return back to smaller values of $\tau$ without passing by singularities. This phenomenon means that the saturation state $\lim_{\tau\to \infty}\widetilde{T}(\tau,b)\to i\epsilon_{\mathcal{P}}/\lambda$ is a terminal attractor of the system, characterizing a physical limit of the evolution of the amplitude with respect to the energy. In fig.\ref{fig:analytical_solutions} we show several solutions $\widetilde{T}(\tau,b)$, running with the variable $\tau$, corresponding to different complex initial conditions, leading to the saturated value $i\epsilon_{\mathcal{P}}/\lambda$. This property unveils the universality of the saturation behavior in high energy independent of the initial fluctuations.

\subsection{Analyticity, dispersion relations, and causality}

In high-energy elastic scattering, the analytic structure of the scattering amplitude reflects fundamental principles of quantum field theory. In particular, causality -- the requirement that no signal propagates faster than light -- implies analyticity of the amplitude in the complex energy plane. These analyticity properties allow for the use of dispersion relations, such as the Kramers–Krönig relation, which link the real and imaginary parts of the amplitude at fixed momentum transfer. 

Here we analyze these aspects for the impact-parameter amplitude $\widetilde{T}(\tau, b)$, where the evolution variable $\tau = \log(s/s_0)$ plays the role of complexified energy. Our starting point is the general solution of the nonlinear evolution equation Eq. (\ref{eq:Analytic_Solution}):
\begin{equation}
\widetilde{T}(\tau, b) = \frac{\widetilde{T}_0(\tau_0, b)}{\left(1 + i\frac{\lambda}{\epsilon_{\mathcal{P}}} \widetilde{T}_0(\tau_0, b)\right) e^{-\epsilon_{\mathcal{P}}(\tau-\tau_0)} - i\frac{\lambda}{\epsilon_{\mathcal{P}}} \widetilde{T}_0(\tau_0, b)}~,
\end{equation}
where $\widetilde{T}_0(\tau_0,b)$ is a complex initial profile at some fixed energy $\tau_0$.

To check for possible singularities, we analyze the denominator:
\begin{equation}
D(\tau) = \left(1 + i\frac{\lambda}{\epsilon_{\mathcal{P}}} \widetilde{T}_0(\tau_0, b)\right) e^{-\epsilon_{\mathcal{P}} (\tau - \tau_0)} - i\frac{\lambda}{\epsilon_{\mathcal{P}}} \widetilde{T}_0(\tau_0, b).
\label{Eq:DenominatorSolution}
\end{equation}

We decompose $\widetilde{T}_0(\tau_0, b)$ into real and imaginary parts:
\[
\widetilde{T}_0 = \widetilde{T}_{0R} + i \widetilde{T}_{0I}~.
\]
Then Eq.~\eqref{Eq:DenominatorSolution} becomes:
\begin{equation}
\left(1 - \frac{\lambda}{\epsilon_{\mathcal{P}}} \widetilde{T}_{0I} + i\frac{\lambda}{\epsilon_{\mathcal{P}}} \widetilde{T}_{0R} \right) e^{-\epsilon_{\mathcal{P}}(\tau - \tau_0)} = i \frac{\lambda}{\epsilon_{\mathcal{P}}} (\widetilde{T}_{0R} + i \widetilde{T}_{0I})~.
\end{equation}

Equating real and imaginary parts, we find:
\begin{align}
\left(1 - \frac{\lambda}{\epsilon_{\mathcal{P}}} \widetilde{T}_{0I} \right) e^{-\epsilon_{\mathcal{P}}(\tau - \tau_0)} &= - \frac{\lambda}{\epsilon_{\mathcal{P}}} \widetilde{T}_{0I}, \label{Eq:Denominator_Real} \\
\frac{\lambda}{\epsilon_{\mathcal{P}}} \widetilde{T}_{0R} \,e^{-\epsilon_{\mathcal{P}}(\tau - \tau_0)} &= \frac{\lambda}{\epsilon_{\mathcal{P}}} \widetilde{T}_{0R}. \label{Eq:Denominator_Imaginary}
\end{align}

Equation~\eqref{Eq:Denominator_Imaginary} implies that if $\widetilde{T}_{0R} \neq 0$, then $e^{-\epsilon_{\mathcal{P}}(\tau - \tau_0)} = 1$, i.e., $\tau = \tau_0$ \footnote{Of course we disregard $\epsilon_{\mathcal{P}}=0$ since this is a trivial solution.}. Plugging this into Eq.~\eqref{Eq:Denominator_Real}, the equation cannot be satisfied unless $\widetilde{T}_{0I} = 0$, which again leads to a trivial case.

If $\widetilde{T}_{0R} = 0$, one finds:
\begin{equation}
\left(1 - \frac{\lambda}{\epsilon_{\mathcal{P}}} \widetilde{T}_{0I} \right) e^{-\epsilon_{\mathcal{P}}(\tau - \tau_0)} = - \frac{\lambda}{\epsilon_{\mathcal{P}}} \widetilde{T}_{0I}~.
\end{equation}
This has no real solution for $\tau > \tau_0$ if $\widetilde{T}_{0I}$ is bounded and $\lambda, \epsilon_{\mathcal{P}} > 0$. Therefore, the solution $\widetilde{T}(\tau, b)$ is analytic in the physical region, with potential singularities only for complex values of $\tau$ outside the physical sheet.

\subsubsection{Causality and dispersion relations.}
Causality in relativistic field theory implies analyticity in the physical (first) sheet of the complex $s$-plane. In particular, the amplitude must be analytic in the upper-half plane of $\sqrt{s}$. Translating this to $\tau = \log s$, one obtains:
\begin{equation}
\text{Im}(\tau) > 0 \quad \Rightarrow \quad \widetilde{T}(\tau, b) \text{ analytic}.
\end{equation}
This justifies the use of standard dispersion relations, such as:
\begin{equation}
\text{Re} \,\widetilde{T}(s,b) = \frac{1}{\pi} \mathcal{P} \int_{-\infty}^{+\infty} \frac{\text{Im} \,\widetilde{T}(s',b)}{s' - s} ds'~,
\end{equation}
which follow from the analyticity properties of causal amplitudes.

To verify this explicitly, one may apply the Cauchy–Riemann conditions. Writing $\tau = \tau_R + i \tau_I$ and $\widetilde{T} = T_R + i T_I$, one obtains:
\begin{eqnarray}
\frac{\partial\, {\rm Re} \,\widetilde{T}(\tau_R,\tau_I)}{\partial \tau_I}=\frac{\partial\, {\rm Im}~\widetilde{T}(\tau_R,\tau_I)    }{\partial \tau_R}    
\end{eqnarray}
and
\begin{eqnarray}
\frac{\partial\, {\rm Re} \,\widetilde{T}(\tau_R,\tau_I)}{\partial \tau_r}=-\frac{\partial \,{\rm Im}~\widetilde{T}(\tau_R,\tau_I)    }{\partial \tau_I}    ~,
\end{eqnarray}
confirming local analyticity in the complex $\tau$-plane.

\subsubsection{Analyticity in $t$ and $b$-space structure.}
The analyticity in momentum transfer $t$ is reflected in the asymptotic behavior of the impact-parameter amplitude. Since our formalism assumes that $\widetilde{T}_0(b)$ falls off exponentially or faster at large $b$, the transformed amplitude $T(s,t)$ will inherit analyticity in $t$ as ensured by the Paley–Wiener theorem. In particular, the decay encodes the presence of a mass gap in the crossed channel, consistent with analytic structure in $t$ near $t=0$.

Although we do not model the $t$-channel singularities explicitly, they are effectively encoded in the profile of $\widetilde{T}_0(b)$ and preserved under the nonlinear evolution. This structure allows us to make phenomenological predictions at the level of differential cross sections without introducing unphysical artifacts.

We conclude that the evolved amplitude $\widetilde{T}(\tau, b)$ is analytic in the physical domain, both in $\tau$ and in $t$ via its $b$-space properties. The absence of singularities ensures that dispersion relations hold and causality is preserved. Thus, the formalism respects fundamental analytic constraints required by relativistic quantum field theory.

\section{Numerical Implementation and Phenomenological Analysis}

Although the present work focuses on the theoretical structure of the complex non-linear evolution equation, its application to phenomenological models is promising.  As an evolution equation our solution requires a complex $b$-profile in a fixed energy, $\widetilde{T}_0(\tau_0,b)$, to evolve the scattering process along rapidity. In this section we briefly describe our results in the Fourier-transformed space in two regimes: the full $t$-range (corresponding to full $b$-profiles) and the forward scattering (corresponding to large $b$-profiles).
Realistic $b$-space profiles extracted from elastic scattering data can serve as initial conditions for a broad $t$-range. Some examples: 
\begin{itemize}
    \item Kohara-Ferreira-Kodama (KFK) model  \cite{kohara:2014} --- based on the stochastic vacuum model \cite{Dosch:1994} uses asymptotic forms of Wilson loop correlators to describe the scattering amplitudes.
    \item Bourrely-Soffer-Wu (BSW) model\cite{BSW:1979} --- an eikonalized model inspired in the crossing symmetry of $s$ and $u$ Mandelstam variables from field theories.
    \item Donnachie-Landshoff (DL) model \cite{Donnachie:2013xia} --- based on single-Pomeron exchange plus double-Pomeron exchange (to account for unitarity) in the $t$ channel in the Regge formalism. 
    \item   Białas-Bzdak model (BB) \cite{Bialas:2006kw} --- based on the modelization of the proton as a pair of quark and di-quark, distributed according to a Gaussian probability density. The model evokes Glauber techniques to superpose the interactions. The model was extended by Csörgő and Nemes to account the real part of the amplitude \cite{Nemes:2012cp,Csorgo:2013usx}. This is the so-called RealBB model.
\end{itemize}

Our preliminary studies show that the evolution of 
 full $t$-profiles captures the main features of the energy dependence of the total cross section and elastic differential cross section.  To capture the Coulomb interference details we use forward amplitudes based on data-driven models. 
A detailed phenomenological analysis, including fitting procedures and predictions for LHC energies and beyond, will be presented in a forthcoming publication.

The models above discussed can be represented in $b$-space, where the real and imaginary parts of their amplitudes can be  decoupled to be injected into our equation. 
The four models described in full $t$-range, have both the $b$ and $s$ dependencies, albeit for our purposes we take each of these models at a single energy, where it has a good match with a specific experimental dataset. The similarities between the four models lie in the behavior of the amplitudes. In momentum space, the real parts fall to zero faster than the imaginary part, crossing zero according to Martin's theorem \cite{Martin:1997}, becoming negative and then reducing again to zero, whereas the imaginary parts fall to zero later constraining the position of the dip. The interplay between real and imaginary parts around the dip gives its magnitude.

The model for the forward scattering is based on exponential forms in $t$ for real and imaginary amplitudes with a linear $t$-term to account for the passage of the real amplitude through zero. This model was described in our previous works \cite{Kohara:2017ats, Kohara:2018wng} with its extension to account the energy dependence \cite{Kohara:2019qoq} and the analytical connection between real and imaginary parts. This model describes accurately the behavior of the real amplitude in the forward regime up to $t\simeq 0.2$ GeV$^2$.

To determine the physical quantities $\lambda$ and $\widetilde{\epsilon}$ adequate for each model, we use the real $\widetilde{T}_{0R}(b)$ and imaginary $\widetilde{T}_{0I}(b)$  profiles in a given energy from a specific model and  we fit our solution in a different energy. To avoid bias, we usually take a large energy gap, for instance, if we give as an input an ISR energy we fit a LHC energy where the distributions were measured in a broad momentum range. Of course different methods of fitting would be possible, but we chose the simplest $\chi^2$ method. The differential cross section is written as the absolute square of our evolution equation given by the solution of Eq.(\ref{eq:Complex_Logistic_2})
\begin{eqnarray}
   \widetilde{T}(\tau,b)=\frac{\widetilde{T}_0(\tau_0,b)}{\Big(1+i\frac{\lambda}{\widetilde{\epsilon}}\widetilde{T}_0(\tau_0,b)\Big)\,e^{-\widetilde{\epsilon}(\tau-\tau_0)}-i\frac{\lambda }{\widetilde{\epsilon}}\widetilde{T}_0(\tau_0,b) }~,
   \label{eq:Analytic_Solution_f}
\end{eqnarray}
with a complex $\widetilde{\epsilon}$ . We anticipate that for long $t$  we consider $\epsilon_{I}=0$, since this quantity is relevant only in the very forward scattering where the real amplitude can be determined with accuracy. Since the long-$t$ models are aimed to describe the global structure of the data they are more insensitive to $\epsilon_{I}$. In the very forward scattering this parameter shows to be relevant to describe the values of $\rho$.

\subsection{From Impact Parameter to Differential Cross Section}

To compute the physical observable $d\sigma/dt$, we perform a Fourier transform of the evolved amplitude in impact parameter space. The scattering amplitude in momentum transfer space is given by (see appendix A)
\begin{equation}
    T(s,t) =  \int_0^\infty db\, b\, J_0(b\sqrt{-t})\, \widetilde{T}(s,b),
    \label{eq:FourierTransform}
\end{equation}
where $J_0$ is the Bessel function of the first kind, and $\widetilde{T}(s,b)$ is the complex amplitude\footnote{To simplify the notation we assume $\widetilde{T}(\tau,b)=\widetilde{T}(s,b)$
} obtained from Eq.~(\ref{eq:Analytic_Solution_f}). The normalization is consistent with the convention used in Regge theory and ensures unitarity when integrated over $t$. 
The differential elastic cross section is then given by
\begin{equation}
    \frac{d\sigma}{dt} = \pi\, |T(s,t)|^2.
    \label{eq:DifferentialCrossSection}
\end{equation}

\subsubsection{Coulomb Interference}

In the very forward region, the electromagnetic -- Coulomb -- interaction plays a non-negligible role and must be taken into account in the amplitude. The full amplitude is composed of the nuclear amplitude $T(s,t)$ and the Coulomb amplitude $T_C(s,t)$ with a relative phase $\phi(s,t)$:
\begin{equation}
    T_{\text{full}}(s,t) = T(s,t) + e^{i\phi(s,t)} T_C(s,t).
    \label{eq:FullAmplitude}
\end{equation}

The Coulomb amplitude for proton-proton scattering, in the one-photon exchange approximation, is given by
\begin{equation}
    T_C(s,t) = \mp \frac{2\alpha}{|t|} F^2(t),
\end{equation}
where $\alpha$ is the fine-structure constant and $F(t)$ is the proton electromagnetic form factor, commonly parametrized by the dipole form
\begin{equation}
    F(t) = \frac{1}{\left(1 - \frac{t}{\Lambda^2}\right)^2}, \quad \Lambda^2 = 0.71\, \text{GeV}^2.
\end{equation}

The relative phase $\phi(s,t)$ is the West–Yennie phase, approximated by
\begin{equation}
    \phi(s,t) = \pm \left[ \ln\left(\frac{B|t|}{2}\right) + \gamma \right],
\end{equation}
where $B$ is the slope parameter of the nuclear amplitude at small $|t|$, and $\gamma$ is the Euler–Mascheroni constant. The sign depends on the particle charges: negative for pp and positive for p\=p scattering.

Finally, the differential cross section including the Coulomb interference is
\begin{equation}
    \frac{d\sigma}{dt} = \pi\, \left| T(s,t) + e^{i\phi(s,t)} T_C(s,t) \right|^2.
    \label{eq:FullDifferentialCrossSection}
\end{equation}

\subsubsection{Practical Implementation}

In practical terms:
 $\widetilde{T}_0(b)$ is taken from a chosen model at energy $\sqrt{s_0}$;
 $\widetilde{T}(b,\tau)$ is computed at new energy $\sqrt{s}$ using Eq.~(\ref{eq:Analytic_Solution_f});
 $T(s,t)$ is obtained by numerical integration of Eq.~(\ref{eq:FourierTransform});
 $d\sigma/dt$ is calculated from Eq.~(\ref{eq:DifferentialCrossSection}) or Eq.~(\ref{eq:FullDifferentialCrossSection}) depending on the $t$ range.

This set of steps allows for the reproduction of results using any input model for the initial amplitude. Parameters $\lambda$ and $\widetilde{\epsilon}$ are fitted only once to match the evolution from $\sqrt{s_0}$ to $\sqrt{s}$, allowing fully predictive interpolation and extrapolation in energy. 

In appendix A we give the kinematical factors, the definition of the invariant amplitudes and its connection with the $T(s,t)$ amplitude as well as the normalizations of the differential and total cross sections used in our work.

\subsection{$t$-dependence of elastic scattering: forward and large-$|t|$ behavior}

In the present work we do not aim to make the full phenomenological study of our equation, but we outline the basic steps to compare it with the data. The determination of the  parameters of the evolution equation is performed in a two-step procedure, reflecting their distinct physical roles.

The parameter $\epsilon_{\mathcal{P}}$, which governs the energy dependence of the scattering amplitude, is determined using a phenomenological profile taken as initial condition at a reference energy ($\sqrt{s} = 52$ GeV for pp).  In practice, $\epsilon_{\mathcal{P}}$ is determined by fitting Eq. (\ref{Eq:Total_Cross}) to the available total cross section data, while keeping $\lambda$ fixed. While this profile fixes the normalization at the initial scale, $\epsilon_{\mathcal{P}}$ controls the evolution with energy.
Its value is obtained by requiring that the evolution equation reproduces the observed energy dependence of the total cross section over a wide range of energies. This procedure ensures that $\epsilon_{\mathcal{P}}$ is constrained by the global energy behavior of the data, rather than by a local fit at a single energy.

The parameter $\lambda$, associated with the nonlinear saturation effects, is constrained using high-energy data. In practice, $\lambda$ is adjusted using Eq.(\ref{eq:FullDifferentialCrossSection}) to reproduce the behavior of the amplitude at LHC energies ($\sqrt{s}$ = 13 TeV for pp), while keeping $\epsilon_{\mathcal{P}}$ fixed. We also tested to determine the parameter $\lambda$ in energies $\sqrt{s}=$ 7 and 8 TeV, and we obtain consitent results. The same procedure applies to p\=p, but due to the lack of experimental data in both energy and momentum transferred-$t$ we are limited to choosing an initial energy scale. In this case, we use the CERN/SPS data at 540 GeV \cite{CERN:540ppbar} as the initial scale and we adjust the parameter $\lambda$ at 53 GeV \cite{Breakstone:1985}.

It is worth emphasizing that the determination of $\epsilon_{\mathcal{P}}$ and $\lambda$ relies on different observables: the total cross section for $\epsilon_{\mathcal{P}}$, and the differential cross section for $\lambda$. This separation naturally leads to their effective independence and justifies the sequential determination procedure adopted in this work.

An important feature of the present framework is that the parameter $\lambda$ is not freely adjustable. Its value is strongly constrained by the requirement of a stable and physically meaningful evolution. In particular, $\lambda$ plays a crucial role in controlling the position of the dip in the differential cross section.

In particular, for a wide class of initial conditions, including different phenomenological parametrizations, only a relatively narrow range of $\lambda$ values leads to acceptable solutions, avoiding unphysical behavior such as oscillations in momentum-transfer space or inconsistencies with experimental data (as can be seen in Table \ref{tab:parameters}).

This indicates that $\lambda$ plays the role of a genuinely dynamical parameter, rather than a simple free fitting parameter, and suggests a certain degree of universality of the nonlinear effects encoded in the evolution equation.

The physical parameters are given in table \ref{tab:parameters}. For full-$t$ models, the imaginary contribution of the mass $\epsilon_I$ seems to be insensitive and we set it to zero. For the forward model used as initial condition, $\epsilon_I$ becomes sensitive and adjusts the real part. As show in table \ref{tab:parameters} the effective mass and coupling are equal for the forward model, leading to further simplification to the evolution equation.

\begin{table}[h]
\centering
\begin{tabular}{cccc}
      \qquad\qquad\qquad                 & $\epsilon_R$& $\lambda$  & $\epsilon_I$  \\ \hline\hline
KFK & 0.125 & 0.098 & -\\ 
BSW  & 0.124 & 0.092 & -\\
DL  & 0.128 & 0.103 & -\\
RealBB  & 0.135 & 0.095 & - \\
\hline
Forward  & 0.15 & 0.15 & 0.014\\
\hline
\end{tabular}
\caption{The parameters $\epsilon_R$, $\epsilon_I$ and $\lambda$ are presented separately for KFK, BSW, DL, RealBB and the Forward models used as initial conditions.
}
\label{tab:parameters}
\end{table}

\begin{figure}[H]
    \centering
\includegraphics[width=0.7\linewidth]{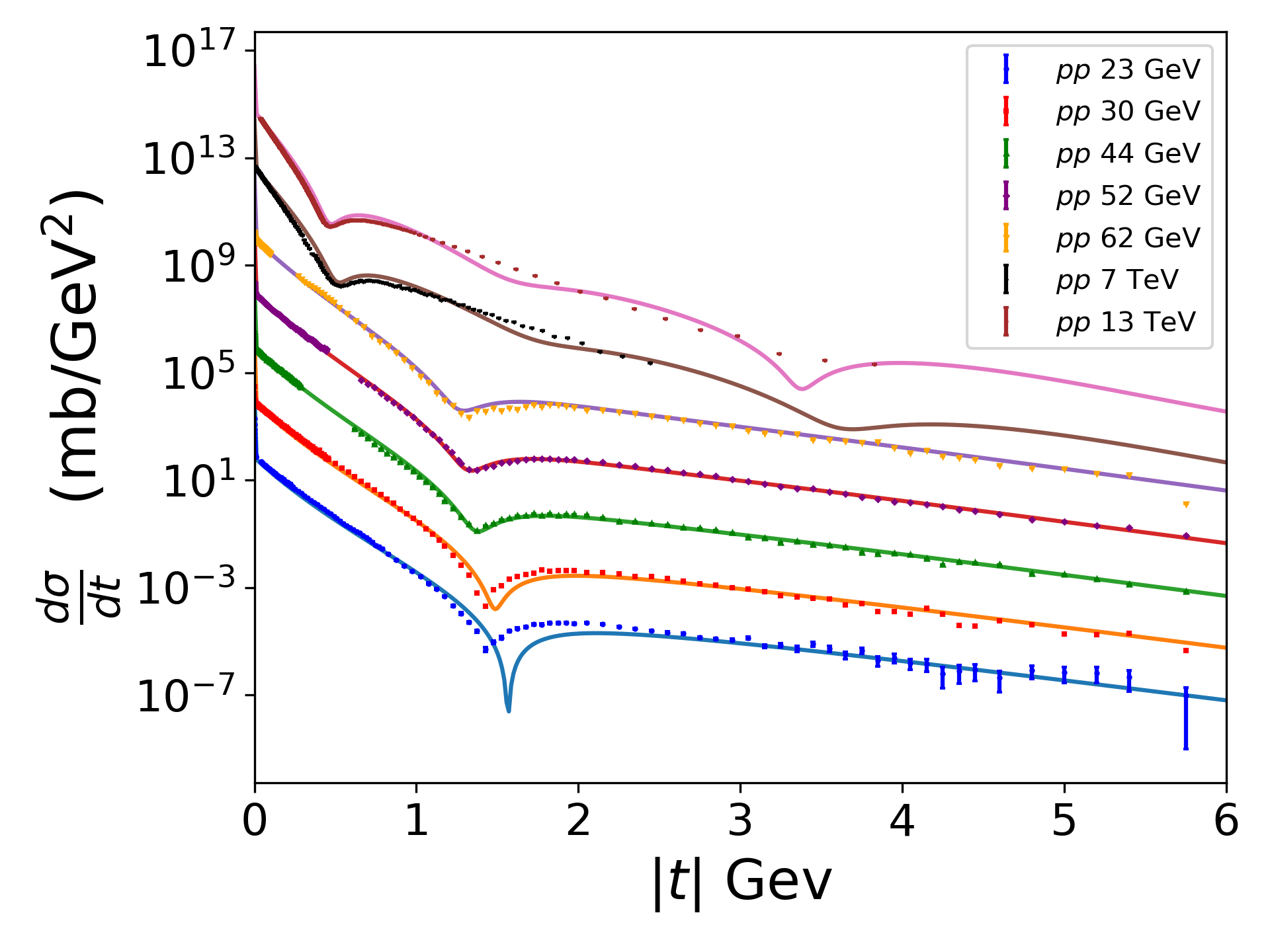}
\includegraphics[width=0.7\linewidth]{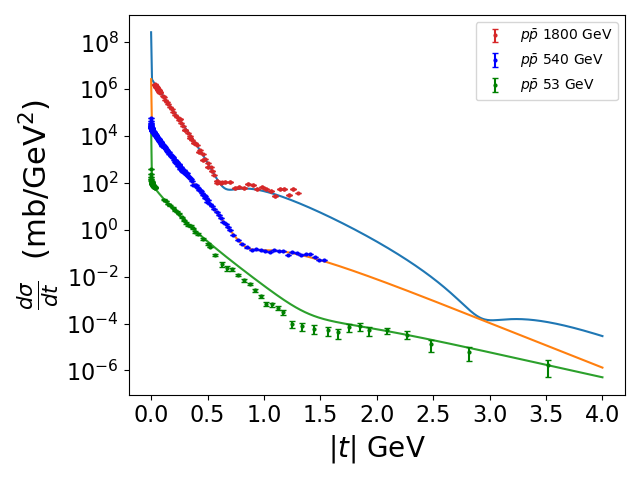}
\caption{The first figure shows the descriptions of the pp differential cross section data from ISR to LHC energies using the KFK model as the initial condition at 52 GeV. It is interesting to observe that below 30 GeV the position of the dip starts to deviate from the data, which could be explained at the low energy range by the Regge trajectories, not taken into account here. The second figure shows the p\=p data from ISR/CERN to TEVATRON/FERMILAB energies. The initial condition is taken at 540 GeV SPS/CERN experiment. The shallow dip structures are remarkable in all the three energies. 
}

    \label{fig:Profiles_IC}
\end{figure}

  In figures \ref{fig:Profiles_IC} we show our preliminary results for the differential cross section of pp and p\=p, using KFK model as  initial condition. In the first figure, we show pp predictions using KFK at 52 GeV as an input energy scale. It is interesting to notice that the evolution equation captures the main features of the differential cross section data in a wide energy range. For lower energies, say below 30 GeV, the devations around the dip could be possible explained by the standard Regge trajectories, not taken into account in the current work. The second figure are the predictions for the p\=p data. The initial energy scale is 540 GeV from SPS/CERN experiment. Since this set of data was given in terms of $dN/dt$ we assume the normalization given in \cite{AKOHARA:2015} to construct the differential cross section data.  We stress that the results shown in Fig. \ref{fig:Profiles_IC} are not obtained from fits to the differential cross section data, but correspond to genuine predictions of the evolution equation once the parameters are fixed. We observe that the structure of the dip and bump are naturally reproduced by the evolution equation from ISR to LHC energies. The other models played as initial conditions give similar description with local changes in different $t$ regions, reflecting the differences from the initial conditions.

\begin{figure}[H]
    \centering
\includegraphics[width=0.7\linewidth]{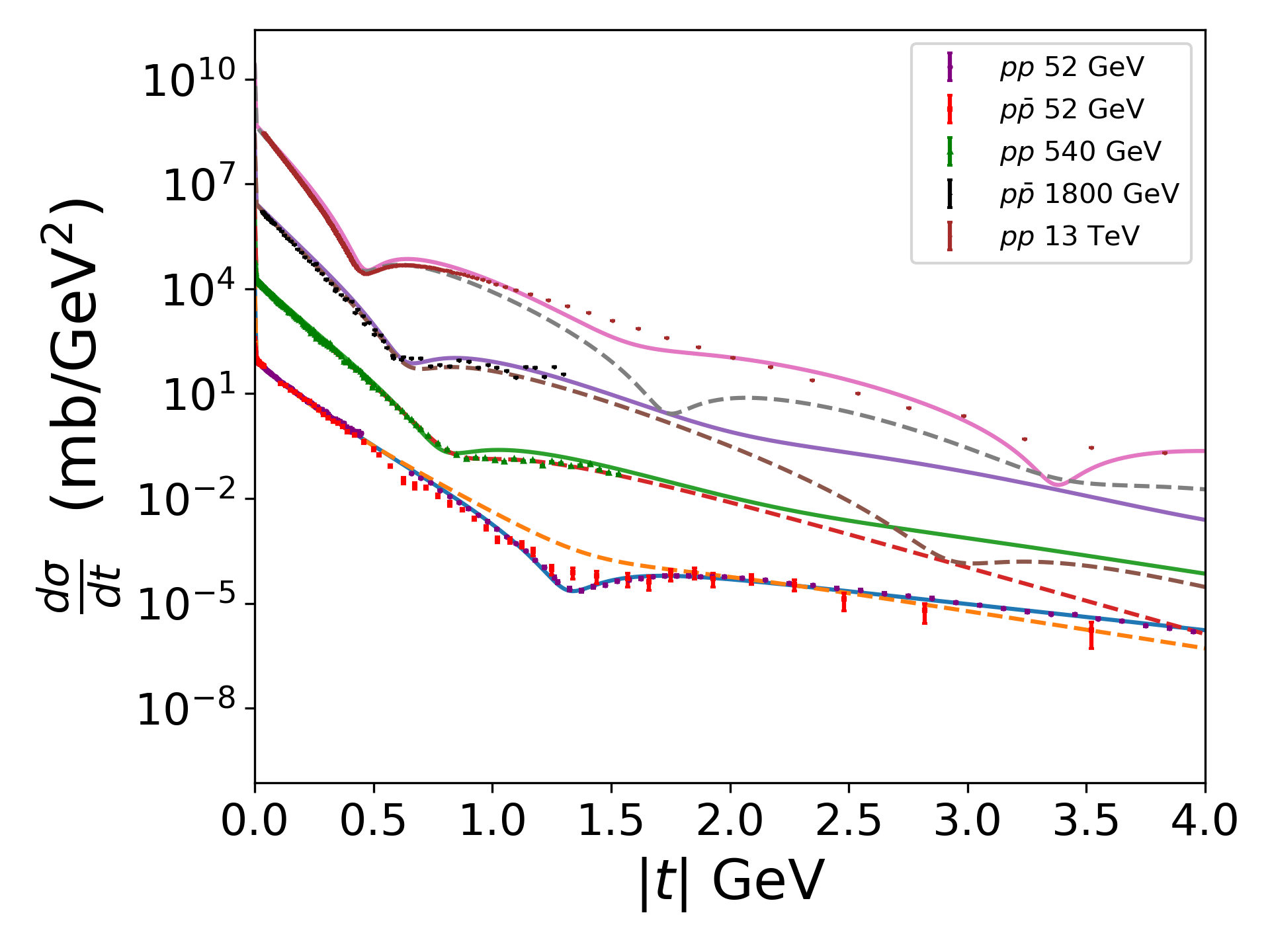}
\caption{We compare pp and p\=p curves from our evolution equation using two different profiles as initial conditions. The dashed curves represent the p\=p solution whereas  the solid line represent pp. It is interesting to note that the profiles for pp and p\=p evolve and the position of the dip follow the same behaviour. At high energies the magnitude of the dips are similar. 
}
\label{Figure:comparision}
\end{figure}

In Fig. \ref{Figure:comparision} we show the comparision between pp and p\=p according to our evolution equation using two different profiles as initial conditions; one for pp and another for p\=p. At 52 GeV we represent both pp and p\=p experimental points with our predictions. While the dashed lines represent p\=p solutions, the solid lines are pp representation of our evolution equation. It is interesting to observe that even when the initial condition are given at different energies and different profiles, the position of the dips do not change with respect to pp and p\=p at the same energy. Moreover, at higher energies --- 13 TeV --- the magnitude of the dip and bump structure are similar. The visible changes happen at large-$t$ domain.

\begin{figure}
    \centering
    \includegraphics[width=.9\linewidth]{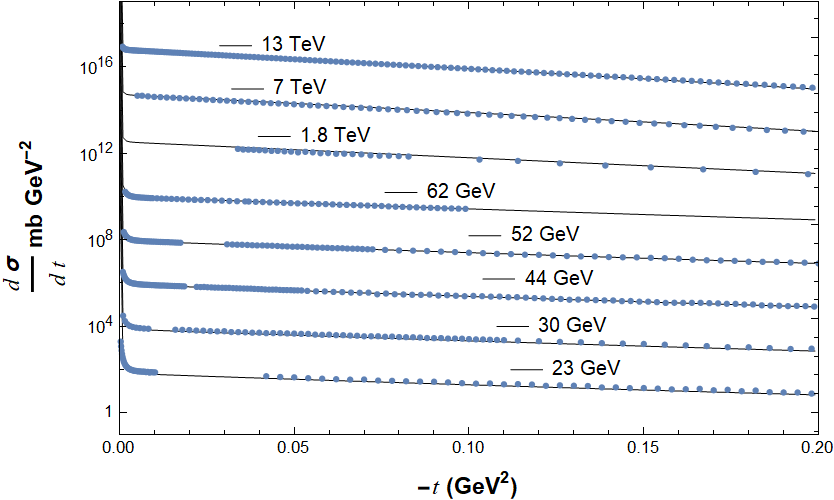}
    \caption{Differential cross sections in the forward region ($0 < |t| < 0.2$ GeV$^2$), obtained from the evolved forward scattering profile with initial conditions at 52.8 GeV. The real part of the amplitude includes the Coulomb contribution. }
    \label{fig:forward_diff}
\end{figure}

\begin{figure}[H]
    \centering
    \includegraphics[width=0.8\linewidth]{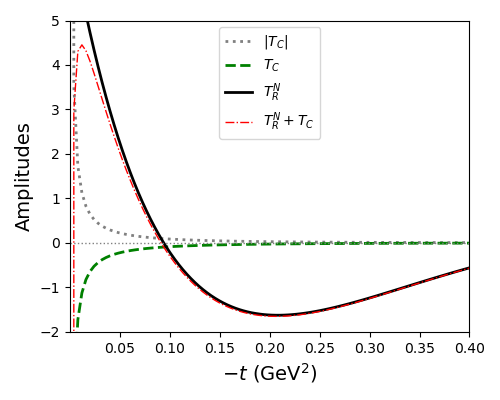}
    \caption{The real nuclear amplitude at 13 TeV with the Coulomb part. The dashed-dotted line represent the sum $T_R^N+T_C$, solid line is $T_R^N$ and the dotted line is the Coulomb amplitude. Note that the nuclear and Coulomb amplitudes interplays creating two zeros near the origin. The first zero is the pure compensation of the nuclear part with the imaginary negative Coulomb interaction and the second zero is the so called Martin's zero. } 
    \label{fig:forward_real_ampl}
\end{figure}

In fig.\ref{fig:forward_diff} we show our results for the differential cross sections in the range ($0<|t|<0.2$ GeV$^{2}$) for the forward profile given as IC at 52.8 GeV. In this figure we add the Coulomb amplitude to the real part. In fig.\ref{fig:forward_real_ampl} we show the real part of the forward amplitude coming from our evolution equation at 13 TeV as an example and we compare it with the Coulomb amplitude, given by the electromagnetic interaction of the charged hadrons. We observe that the real part of the amplitude crosses zero twice. The first zero arises from the interference with the Coulomb amplitude, which is negative in the case of pp scattering. The second zero, known as the Martin zero, originates purely from the nuclear interaction. Qualitatively, the behavior of the amplitude in the very forward region resembles that of the large-$t$ models; however, in those models, the Martin zero typically appears at larger values of $|t|$, shifted further to the right in momentum transfer.

\subsection{Total cross section $\sigma_{\rm tot.}$}

\begin{figure}[H]
    \centering
    \includegraphics[width=0.8\linewidth]{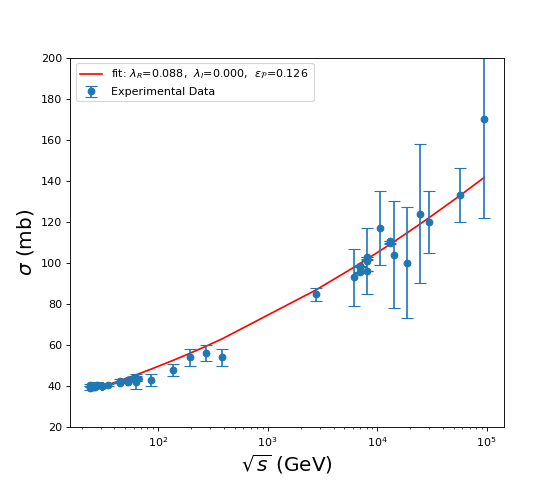}
    \caption{Fit of the experimental data from PDG (COMPETE) compilation \cite{Workman:2022ynf} from ISR to cosmic ray energies. }
    \label{fig:total_cross}
\end{figure}

The imaginary part of Eq.(\ref{eq:Analytic_Solution_f}) into Eq.(\ref{Eq:Total_Cross}) determines the total cross section up to two physical quantities: $\widetilde{\epsilon}$ and $\lambda$. We notice that the total cross section is mainly influenced by the effective real mass $\epsilon_R$, while the other quantities do not affect qualitatively the energy dependence of $\sigma_{\rm tot.}$. For  large-$t$ models as initial conditions the effective real mass is $\epsilon_R\sim 0.12$, describing accurately the experimental data from ISR to LHC energies and being extrapolated to the very high energy cosmic rays.  In fig.(\ref{fig:total_cross}) we represent our solution using the KFK model as an example to describe the total cross section data. The other models lead to similar description.

\subsection{The ratio $\rho$}
The ratio $\rho$ is a physical parameter determined at the origin. In most of the experimental analyses, this quantity comes from simplified exponential forms where the real and imaginary amplitudes have the same $t$ dependence. It is important to stress that this procedure of extracting $\rho$ is already an assumption and as a consequence it is model dependent.

The dispersion relations connect the real and imaginary parts of amplitudes. Since the total cross section is given by the pure imaginary part at the origin, its behavior, up to asymptotic energies could give us the value of the parameter $\rho$. Many studies have been conducted to predict $\rho$ from the parametrization of the total cross section \cite{Ferreira:2018ijj,Fagundes:2017iwb, Avila:2003cu, Avila:2005rg,Workman:2022ynf}. It is important to stress that in order to calculate $\rho$ from dispersion relations we need the knowledge of the entire $\sigma_{\rm tot.}$ which for very high energies may be not realistic. Extrapolations are always used and dispersion relations are not  universal  in this sense. Since our equation connects analytically the real and the imaginary parts in the entire $s$ domain, the dispersion relations constraints are automatically satisfied.

For the long-$t$ initial conditions we observe that  $\rho$ is almost constant in a broad energy range  with an asymptotic decreasing trend and from Eq.(\ref{Eq:Solution_Asymptotic}) when $\tau\to \infty$ and $\epsilon_I=0$ the amplitude is pure imaginary and $\rho\to0$. However, for forward scattering initial conditions, which is based on a data-driven model, the trend of $\rho$ approaches the experimental behavior and the imaginary part of the effective mass $\epsilon_I$ plays an important role. 

Our conclusion is that $\rho$ should not be treated as an experimental value since it is hidden behind the imaginary part, which is ten times larger than the real part at the origin, and a slightly change in the parametrization of the amplitudes could lead to strong modifications in the numerical values.

As signalized before, other interesting interference effects on the real part could be potentially observed at LHC energies \cite{Kohara:2022ktp, Selyugin:2023txo}, constraining the behavior of the real amplitude showing a clear interference pattern. If this is the case, as a natural consequence, the values of $\rho$ would be more realistic.

\subsection{Crossing symmetry}

 It is noteworthy that our evolution equation, without introducing any explicit dynamical distinction between pp and p\=p scattering, is able to reproduce the main qualitative features observed in the experimental data over a wide energy range, starting from initial conditions  to ISR pp data at 52 GeV and SPS p\=p data at 540 GeV, as shown in Fig.~\ref{Figure:comparision}.

Within our framework, both pp and p\=p profile amplitudes are evolved using the same nonlinear equation. No channel-dependent ingredients are introduced in the evolution dynamics. As a consequence, any differences between pp and p\=p observables at higher energies originate solely from the initial conditions at the ISR and SPS scales and from the nonlinear nature of the evolution.

In particular, the evolution leads to a non-trivial modification of the dip-bump structure in the differential cross section as energy increases. At intermediate energies, such as 540 and 1800 GeV, the dip appears less pronounced, while at LHC energies it becomes sharper again. This behaviour emerges dynamically from the interplay between nonlinear saturation effects in impact-parameter space and the spatial structure of the initial profiles, rather than from any explicit energy-dependent modification of the model.

This suggests that the observed differences between pp and p\=p differential cross sections in the Tevatron energy range may be understood, at least qualitatively, within the same universal evolution dynamics, without invoking additional dynamical mechanisms beyond those already encoded in the nonlinear equation and the initial conditions.

A more stringent test of this picture would require reconstructing both pp and p\=p amplitudes at ISR energies using a fully data-driven parameterisation, and evolving them independently to higher energies. This would allow a quantitative assessment of whether the observed differences at intermediate energies are fully generated by the evolution, or whether additional charge-conjugation odd contributions (e.g. an Odderon-like component) are required.

\section{Comments and Conclusions}

In this work, we derive a universal evolution equation for elastic scattering of hadrons, rooted in RFT and structurally similar to QCD small-$x$ evolution equations.
Although the nonlinear evolution equation employed here is not derived
directly from small-$x$ QCD, its structure shares qualitative features with
nonlinear evolution equations such as the BK equation, in particular the
presence of a linear growth term and a nonlinear saturation mechanism.
In the present work we do not attempt to establish a quantitative mapping
between the two frameworks. Rather, the connection should be understood
at a phenomenological level, where the nonlinear dynamics in impact-parameter
space plays a role analogous to gluon saturation in small-$x$ QCD. Future work may clarify whether a more quantitative correspondence can be established in specific kinematic regimes

Our formalism reveals that the unitarity constraint naturally emerges from a logistic-type mechanism in rapidity space. Unlike traditional perturbative QCD approaches that rely heavily on transverse diffusion, we demonstrate that in the elastic regime, the diffusive term becomes redundant. Its role is effectively taken over by the intrinsic non-linearity of the strong interaction. As a result, our formalism respects key high-energy constraints, including unitarity, the Froissart bound, and dispersion relations.

Although the evolution equation is formally local in the impact parameter 
$b$, as the underlying Regge-inspired Lagrangian includes rapidity ($\tau=\log s$) derivatives but no spatial gradients, the structure of the amplitude 
$\widetilde{T}(\tau,b)$ reflects a non-trivial interplay between its linear and nonlinear components, both explicitly dependent on $b$. This dependence effectively couples different regions in 
$b$-space through their shared evolution dynamics, inducing an emergent non-locality. As a result, the amplitude evolves towards a saturation regime where each 
$b$-dependent component asymptotically approaches the unitarity (black disk) limit, but the full profile retains a spatially correlated structure.

This framework not only enables predictions from existing input profiles but also serves as a diagnostic tool: if a given model leads to unphysical behavior or incorrect predictions under energy evolution, this signals problems in its $b$-space structure (e.g., lack of a mass gap, excessive localization). Thus, the evolution equation can help constrain the class of phenomenological models compatible with nonperturbative QCD dynamics.

Importantly, we find that the real part of the Pomeron mass (Pomeron intercept) governs the total cross section, while the non-linear coupling $\lambda$ controls the energy evolution of dip and bump structures in the differential cross section. Our results naturally explain the transition from pronounced dips at ISR energies to shallower structures at higher energies, such as in p$\bar{\text{p}}$ scattering at 540 and 1800 GeV.

Regarding the forward scattering regime, we argue that the $\rho$ parameter (the ratio of real to imaginary parts at $t=0$) should not be treated as a direct observable. Instead, it is more robust to analyze features of the differential cross section that signal nontrivial interference between real and imaginary parts. The existence of such structures can place meaningful constraints on the real part of the amplitude, providing a more solid foundation for modeling initial conditions.

The logistic-type nonlinear term leads to a dynamical saturation of the scattering amplitude, ensuring unitarity and preventing unbounded growth with energy. This results in an effective suppression of large impact-parameter contributions, thereby shaping the transverse profile of the interaction region. This behavior reflects the role of nonlinear unitarized evolution in impact-parameter space in controlling the spatial structure of the scattering amplitude.

Finally, 
the evolution equation admits a natural interpretation within the framework of dynamical systems theory. The nonlinear saturation term, analogous to a damping force, ensures the system's convergence toward a finite asymptotic amplitude, respecting unitarity. The equation features two fixed points: a trivial one at zero amplitude and a non-trivial saturated solution, proportional to 
$i\,\epsilon_{\mathcal{P}}/\lambda$, which represents the unitarity limit. A linear stability analysis reveals that the zero-amplitude fixed point is unstable, while the saturated state is stable, confirming that the system dynamically evolves toward saturation. This behavior mirrors classical damped systems, where dissipation drives the system to equilibrium. Moreover, the structure of the equation aligns with reaction-diffusion models such as FKPP, suggesting that elastic scattering evolution may belong to a broader universality class of nonlinear systems characterized by saturation and wavefront propagation. These dynamical features strengthen the robustness and predictive power of our framework, offering a physically transparent mechanism for unitarization.

\appendix

\section{Kinematic factors and amplitude conventions  \label{AppendixA}}

We consider $2\to 2$ elastic scattering of equal-mass hadrons ($pp$ or $p\bar{p}$) in the high-energy limit ($s \gg m^2$). The $S$-matrix is written in momentum space as
\begin{equation}
S_{fi} = \delta_{fi} + i (2\pi)^4 \delta^{(4)}(p_f - p_i) \, \mathcal{A}_{fi},
\end{equation}
$\mathcal{A}(s,t)$ is the invariant amplitude. In the center of mass frame, the differential cross-section reads
\begin{equation}
\frac{d\sigma}{dt} = \frac{1}{64\pi s\, p_{\mathrm{cm}}^2} \, |\mathcal{A}(s,t)|^2,
\end{equation}
with $p_{\mathrm{cm}} = \sqrt{s/4 - m^2} \simeq \sqrt{s}/2$ at high energies. In this limit,
\begin{equation}
\frac{d\sigma}{dt} \approx \frac{1}{16\pi s^2} \, |\mathcal{A}(s,t)|^2.
\end{equation}

It is convenient to factor out the leading $s$ dependence by defining the reduced amplitude $T(s,t)$ via
\begin{equation}
 i\, T(s,t)=\frac{\mathcal{A}(s,t) }{4\pi s},
\end{equation}
so that
\begin{equation}
\frac{d\sigma}{dt} = \,\pi\, |T(s,t)|^2.
\end{equation}

The impact-parameter amplitude $\widetilde{T}(s,b)$ is then defined by the Fourier--Bessel transform
\begin{equation}
T(s,t) = \int_0^\infty b\, db \, J_0(b\sqrt{-t}) \, \widetilde{T}(s,b),
\end{equation}
with inverse
\begin{equation}
\widetilde{T}(s,b) = \int_0^\infty q\, dq \, J_0(b q) \, T(s,t=-q^2).
\end{equation}

The optical theorem in impact-parameter space follows as
\begin{equation}
\sigma_{\mathrm{tot}}(s) = \frac{\mathrm{Im}\,\mathcal{A}(s,0)}{s}  = 4\pi \int_0^\infty b \, db \, \mathrm{Im}\,T(s,b).
\end{equation}

\section{Deriving the evolution equation from RFT \label{Appendix_B}}

\subsection{Scope and assumptions}

We work in the framework of Reggeon Field Theory (RFT) for elastic hadron scattering at high energies ($s \gg m^2$). 
The aim is to derive an effective nonlinear evolution equation for the impact-parameter amplitude $\widetilde{T}(s,b)$ starting from an RFT Lagrangian with triple-Pomeron interactions. 
The derivation presented here follows the operator formalism of Alessandrini, Amati, and Ciafaloni~\cite{Alessandrini:1977}, but with modifications suited to our phenomenological goals.  
The main assumptions are:
\begin{itemize}
    \item One-channel elastic scattering of equal-mass particles;
    \item Evolution variable $\tau = \ln(s/s_0)$;
    \item Pomeron intercept $\alpha_{\mathcal{P}}(0) = 1 + \epsilon_{\mathcal{P}}$;
    \item Diffusionless limit $\alpha'=0$ to isolate nonlinear effects;
    \item Semi-classical truncation, neglecting higher-order correlators.
\end{itemize}

\subsection{Brief historical context}

RFT extends Regge phenomenology into a quantum field-theoretic setting, treating Reggeons as quantum fields. 
It was pioneered by Gribov in his seminal work on Reggeon Calculus~\cite{Reggeon-Calculus:1967}, and developed in the 1970s by Alessandrini, Amati, Ciafaloni and collaborators~\cite{Alessandrini:1977,abarbanel1975reggeon,Amati:1976,Allesandrini:1976}. 
In this approach, the Pomeron is the dominant Reggeon at high energies, and its field plays the role of the fundamental degree of freedom. 
The RFT Lagrangian incorporates Reggeon propagation, Reggeon–Reggeon interactions (notably triple-Pomeron vertices), and respects the analytic structure of scattering amplitudes in the complex angular momentum plane.

\subsection{Starting Lagrangian and effective form}

We start from the $2+1$ dimensional RFT Lagrangian density proposed in the 1970's:
\begin{equation}
\mathcal{L} = -\frac{1}{2} \overline{\Psi} \overleftrightarrow{\partial}_{\tau} \,\Psi
-\alpha'\nabla_b \overline{\Psi}\cdot\nabla_b \Psi
+\epsilon_{\mathcal{P}}\,\overline{\Psi}\,\Psi
-i\lambda\,\overline{\Psi}\,(\Psi+\overline{\Psi})\,\Psi \,,
    \label{Eq:Lagrangian_density_RFT_0}
\end{equation}
where $\Psi$ and $\overline{\Psi}$ are the Pomeron field and its conjugate, $\alpha'$ is the Pomeron slope, $\epsilon_{\mathcal{P}}$ the intercept, and $\lambda$ the triple-Pomeron coupling. 
The interaction term has a negative sign dictated by the hybrid Feynman rules of RFT~\cite{abarbanel1975reggeon}, and the vertex is purely imaginary.

In this work, we introduce two phenomenologically motivated modifications:
\begin{enumerate}
  \item In order to isolate the nonlinear dynamics responsible for saturation effects, we consider the diffusionless limit $\alpha'\nabla_b^2$, which reduces the transverse dynamics to a local evolution in impact parameter. This corresponds to focusing on the dominant nonlinear interaction regime while neglecting subleading transverse diffusion effects. In addition with $\alpha'\to0$ we avoid oscillatory artifacts in the Fourier-transformed amplitude.

    \item At the effective level, and motivated by the requirement of stable nonlinear evolution for the elastic amplitude, we consider a modified interaction term that preserves the polynomial structure of the RFT Hamiltonian while allowing bounded solutions for the one-point function.
\end{enumerate}

The resulting $0+1$ dimensional effective Lagrangian reads:
\begin{equation}
\mathcal{L}_{\rm eff.} = -\frac{1}{2} \overline{\Psi} \overleftrightarrow{\partial}_{\tau} \,\Psi
+\epsilon_{\mathcal{P}}\,\overline{\Psi}\,\Psi
+i\lambda\,\overline{\Psi}\,(\Psi+\overline{\Psi})\,\Psi \,.
    \label{Eq:Lagrangian_density_RFT_1}
\end{equation}
Rescaling the fields $q=i\overline{\Psi}$, $p=i\Psi$ makes the Lagrangian purely real:
\begin{equation}
\mathcal{L}_{\rm eff.} = \frac{1}{2} q \overleftrightarrow{\partial}_{\tau} \,p
-\epsilon_{\mathcal{P}}\,q\,p
-\lambda\,q\,(p+q)\,p \,,
    \label{Eq:Lagrangian_density_RFT}
\end{equation}
with $q$ and $p$ depending on $b$ and $\tau$.

\subsection{ From Lagrangian to Hamiltonian}

The Hamiltonian density follows from
\begin{equation}
\mathcal{H}=\sum_{i}\pi_i\,\dot{Q}_i-\mathcal{L}_{\rm eff.}, \qquad
\pi_i=\frac{\partial \mathcal{L}_{\rm eff.}}{\partial \dot{Q}_i}~,
    \label{Eq:momentum}
\end{equation}
with $Q_i = q, p$. Integrating over $\vec{b}$ gives:
\begin{equation}
H(\tau)=\int d^2\vec{b}\,\Big(\epsilon_{\mathcal{P}}\,q\,p+\lambda\,q\,(p+q)\,p\Big).
    \label{Eq:Hamiltonian_RFT}
\end{equation}

Upon canonical quantization, $q$ and $p$ become creation and annihilation operators obeying:
\begin{equation}
[\hat{p}(\vec{b},\tau),\hat{q}(\vec{b}',\tau)]=-\delta^{(2)}(\vec{b}-\vec{b}').
    \label{Eq:Commutator}
\end{equation}

\subsection{Coherent-state formalism and correlation hierarchy}

The imaginary-time Schrödinger equation reads:
\begin{equation}
\frac{\partial  }{\partial \tau}|\psi\rangle = -H|\psi\rangle,
    \label{Eq:Schrodinger}
\end{equation}
where the state $|\psi(\tau)\rangle$ is expanded in terms of an infinite set of correlation functions $G_k(\vec{b}_1,\ldots,\vec{b}_k;\tau)$ through a generalized coherent state:
\begin{equation}
|\psi(\tau)\rangle = e^{-\hat{M}(\tau)}|\psi(0)\rangle,
\label{Eq:Generalized_Coherent_State}
\end{equation}
where the evolution operator $\hat{M}(\tau)$ is 
\begin{eqnarray}
    \hat{M}(\tau)=\sum_{k=1}^{\infty}\frac{1}{k!}\int d^2\vec{b}_1...d^2\vec{b}_k\,\hat{q}(\vec{b}_1)...\hat{q}(\vec{b}_k)\,G_k\Big(\vec{b}_1,...,\vec{b}_k;\tau\Big)~,
    \label{Eq:M_Operator}
\end{eqnarray}
and $|\psi(0)\rangle$ is the perturbative vacuum state of the theory such that 
\begin{eqnarray}
\hat{p}(b_k)|\psi(0)\rangle=0    
\end{eqnarray}
 for any $k$. Note that the n-point correlation function is
 \begin{eqnarray}
G_k\Big(\vec{b}_1,...,\vec{b}_k;\tau\Big)=\langle \psi(0)|\mathcal{T}\Big\{q(\vec{x}_1)...q(\vec{x}_k)p(\vec{y}_1)...p(\vec{y}_k)\Big\}|\psi(0)\rangle ~,
\end{eqnarray}
where the impact parameter is defined as $\vec{b}_k=\vec{x}_k-\vec{y}_k$. 
 
Expanding Eq.(\ref{Eq:Generalized_Coherent_State}) we obtain
\begin{eqnarray}
&&|\psi(\tau)\rangle=\Big(1-\hat{M}(\tau)+\frac{1}{2!}\hat{M}^2(\tau)-\frac{1}{3!}\hat{M}^3(\tau)...\Big)|\psi(0)\rangle  \\
&&=\Big\{1-\Big(\int_b\hat{q}(\vec{b})\,G_1(\vec{b};\tau) +\frac{1}{2!}\int_{b_1,b_2}\hat{q}(\vec{b}_1)\,\hat{q}(\vec{b}_2)\,G_2(\vec{b}_1,\vec{b}_2;\tau)+...\Big)+ \nonumber \\
&&+\frac{1}{2!}\Big(\int_b\hat{q}(\vec{b})\,G_1(\vec{b};\tau) +\frac{1}{2!}\int_{b_1,b_2}\hat{q}(\vec{b}_1)\,\hat{q}(\vec{b}_2)\,G_2(\vec{b}_1,\vec{b}_2;\tau)+...\Big)^2+\nonumber \\
&&-\frac{1}{3!}\Big(\int_b\hat{q}(\vec{b})\,G_1(\vec{b};\tau) +\frac{1}{2!}\int_{b_1,b_2}\hat{q}(\vec{b}_1)\,\hat{q}(\vec{b}_2)\,G_2(\vec{b}_1,\vec{b}_2;\tau)+...\Big)^3+...\Big\}|\psi(0)\rangle ~. \nonumber
    \label{Eq:Expansion_Coherent_State}
\end{eqnarray}
The lowering and raising operators in principle depend on $b$ and $\tau$, but in what follows we disregard the $\tau$ dependence.
The idea is to replace Eq.(\ref{Eq:Generalized_Coherent_State}) in Eq.(\ref{Eq:Schrodinger}) and regroup all the terms with the same order in $\hat{q}$. The left-hand side of Schrödinger equation leads to
\begin{eqnarray}
&&\frac{\partial}{\partial \tau}|\psi(\tau)\rangle=\Big\{-\int_b\hat{q}(\vec{b})\,\frac{\partial}{\partial \tau}G_1(\vec{b};\tau) \nonumber \\
&&-\frac{1}{2!}\int_{b_1,b_2}\hat{q}(\vec{b}_1)\,\hat{q}(\vec{b}_2)\,\Big[\frac{\partial}{\partial \tau}G_2(\vec{b}_1,\vec{b}_2;\tau)-\frac{\partial}{\partial \tau}\Big(G_1(\vec{b}_1;\tau)G_1(\vec{b}_2;\tau)\Big)\Big]+ \nonumber \\
&&-\frac{1}{3!}\int_{b_1,b_2,b_3}\hat{q}(\vec{b}_1)\,\hat{q}(\vec{b}_2)\,\hat{q}(\vec{b}_3)\,\Big[\frac{\partial}{\partial \tau}G_3(\vec{b}_1,\vec{b}_2,\vec{b}_2;\tau) -\frac{3}{2!}\frac{\partial}{\partial \tau}\Big(G_1(\vec{b}_1;\tau)G_2(\vec{b}_2,\vec{b}_3;\tau)\Big)  \nonumber \\
&&-\frac{3}{2!}\frac{\partial}{\partial \tau}\Big(G_1(\vec{b}_3;\tau)G_2(\vec{b}_1,\vec{b}_2;\tau)\Big)+\frac{\partial}{\partial \tau}\Big(G_1(\vec{b}_1;\tau)G_1(\vec{b}_2;\tau)G_1(\vec{b}_3;\tau)\Big)\Big]+...\Big\}~.
    \label{Eq:LHS_Schrodinger}
\end{eqnarray}

The first term in the Hamiltonian on the right-hand side of Eq.(\ref{Eq:Schrodinger}) gives
\begin{eqnarray}
&&-\epsilon_{\mathcal{P}}\Big\{\int_b\hat{q}(b)\,G_1(\vec{b};\tau)+\int_{b_1,b_2}\hat{q}(b_1)\hat{q}(b_2)\Big[G_2(\vec{b}_2,\vec{b}_1;\tau)-G_1(\vec{b}_1;\tau)G_1(\vec{b}_2;\tau)\Big]+\nonumber \\
&&+\frac{1}{3!}\int_{b_1,b_2,b_3}\hat{q}(b_1)\hat{q}(b_2)\hat{q}(b_3)\Big[...\Big]+...\Big\}
    \label{Eq:RHS_first_term}
\end{eqnarray}
The interaction part leads to
\begin{eqnarray}
&&\lambda\Big\{-\int_b\hat{q}(b)\,\Big[G_2(\vec{b},\vec{b};\tau)-G_1(\vec{b};\tau)G_1(\vec{b};\tau)\Big] \\
&&+\int_{b_1,b_2}\hat{q}(b_1)\hat{q}(b_2)\Big[G_1(\vec{b}_1)\delta(\vec{b}_1-\vec{b}_2)+G_3(\vec{b}_2,\vec{b}_1,\vec{b}_1)-2G_1(\vec{b}_1)G_2(\vec{b}_1,\vec{b}_2)+\nonumber \\
&&-G_1(\vec{b}_2)G_2(\vec{b}_1,\vec{b}_1)+G_1(\vec{b}_2)G_1(\vec{b}_1)G_1(\vec{b}_1)\Big]+\frac{1}{3!}\int_{b_1,b_2,b_3}\hat{q}(b_1)\hat{q}(b_2)\hat{q}(b_3)\Big[...\Big]+...\Big\} \nonumber
    \label{Eq:RHS_second_term}
\end{eqnarray}
Regrouping order-by-order leads to an infinity set of coupled differential equations.
To derive the above equations the commutator $(\ref{Eq:Commutator})$ was used.  The first two coupled differential equations are
\begin{eqnarray}
\frac{\partial}{\partial\tau}G_1(\vec{b};\tau)=\epsilon_{\mathcal{P}}\,\Big[1+\frac{\lambda}{\epsilon_{\mathcal{P}}}G_1(\vec{b};\tau)\Big]G_1(\vec{b};\tau)-\lambda \,G_2(\vec{b},\vec{b};\tau)
    \label{Eq:First_Equation}
\end{eqnarray}
and
\begin{eqnarray}
&&\frac{\partial}{\partial\tau}G_2(\vec{b}_1,\vec{b}_2;\tau)=2\epsilon_{\mathcal{P}}\,G_2(\vec{b}_1,\vec{b}_2;\tau)+4\,\lambda G_1(\vec{b}_1;\tau)G_2(\vec{b}_1,\vec{b}_2;\tau)\nonumber \\
&&-2\,\lambda\,G_1(\vec{b}_1;\tau)\,\delta(\vec{b}_1-\vec{b}_2)-2\,\lambda\,G_3(\vec{b}_1,\vec{b}_1,\vec{b}_2)
    \label{Eq:Second_Equation}
\end{eqnarray}

This structure is analogous to the Balitsky hierarchy in small-$x$ QCD~\cite{Balitsky:1995ub,JalilianMarian:1997jx,Weigert:2000gi}.

\subsection{Semi-classical truncation and identification with $T(b,\tau)$}

Neglecting higher-order correlators corresponds to a semi-classical (mean-field) truncation of the Reggeon hierarchy, analogous in spirit to the Balitsky–Kovchegov approximation in small-x QCD. In this approximation, the nonlinear evolution of the one-point function effectively resums multi-Reggeon interactions in a closed form. We obtain
\begin{equation}
\frac{\partial}{\partial\tau}G_1(\vec{b};\tau)=\epsilon_{\mathcal{P}}\left[1+\frac{\lambda}{\epsilon_{\mathcal{P}}}G_1(\vec{b};\tau)\right]G_1(\vec{b};\tau).
    \label{Eq:First_Equation_2}
\end{equation}
Identifying
\begin{equation}
G_1(\vec{b},\tau)\sim i\,\widetilde{T}(b,\tau),
\end{equation}
with complex $\widetilde{T}$, yields the complex logistic equation (Eq.~(\ref{eq:Complex_Logistic}) in the main text). 

The resulting complex logistic equation should therefore be understood as an effective nonlinear evolution equation emerging from Reggeon interactions under controlled approximations, rather than as a fundamental microscopic theory.

\section{Integrals for total cross section for pure imaginary amplitudes \label{Appendix_D}}
To calculate the forward imaginary amplitude from a purely Gaussian imaginary profile we have
\begin{eqnarray}
    K_1=\int b\,db\frac{B\,e^{-\beta\,b^2}}{C+D\,e^{-\beta\,b^2}}=-\frac{B}{2\,D\,\beta}\log(C+D\,e^{-\beta\,b^2}) ~.\label{Eq:Use_Integral_K1_2} 
\end{eqnarray}

The forward imaginary amplitude for a purely exponential profile is given by the integral

\begin{eqnarray}
    K_2=\int b\,db\frac{B\,e^{-\beta\,b}}{C+D\,e^{-\beta\,b}}=-\frac{1}{\beta^2}\Bigg( {\rm Li}_2\Big(\frac{D\,e^{-\beta\,b}}{C}+1\Big)+\log\Big(-\frac{D}{C}\Big)\log\Big(C+D\,e^{-\beta\,b}\Big)\Bigg)~.
    \label{Eq:Use_Integral_K2_3} 
\end{eqnarray}

\subsection*{Integrals of the real and imaginary parts of the amplitudes at $t=0$ for Gaussian profiles }
To calculate the forward real and imaginary amplitudes in $t$ space we need the integrals

\begin{eqnarray}
    I_1=\int b\,db\,\frac{A\,e^{-\beta\,b^2}}{(B\,e^{-\beta\,b^2}+D)^2+C}=-\frac{A}{2\,B\,\sqrt{C}\,b}\tan^{-1}\Big(\frac{B\,e^{-\beta\,b^2}+D}{\sqrt{C}}\Big)
    \label{Eq:Use_Integral_1_2}
\end{eqnarray}

The second kind of integral is 
\begin{eqnarray}
  && I_2=\int b\,db\,\frac{A}{(B\,e^{-\beta\,b^2}+D)^2+C} \\
   &&=\frac{A}{2\,\beta\,((B\,e^{-\beta\,b^2}+D)^2+C)}\Bigg(\frac{D}{\sqrt{C}}\tan^{-1}\Big(\frac{B\,e^{-\beta\,b^2}+D}{\sqrt{C}}\Big)-\log\frac{(B\,e^{-\beta\,b^2})}{(C+(B\,e^{-\beta\,b^2}+D)^2)^{\frac{1}{2}}}\Bigg) \nonumber \label{Eq:Use_Integral_2_3}
\end{eqnarray}

Separating the real and imaginary parts of Eq. (\ref{eq:Analytic_Solution}), and applying Eq. (\ref{eq:Complex_Profile_Gaussian}) as initial condition, we calculate
\begin{eqnarray}
   &&T_{R}(\tau,t=0)= \int_0^{\infty} b\,db\,\,\widetilde{T}_{0R}(\tau,b)\nonumber \\
   &&=\int_0^{\infty} b\,db\,\frac{\widetilde{T}_{0R}(b)\Big[\Big(1-\frac{\lambda}{\epsilon_{\mathcal{P}}}\widetilde{T}_{0I}(b)\Big)\,e^{-\epsilon_{\mathcal{P}}\,\Delta\tau}+\frac{\lambda}{\epsilon}_{\mathcal{P}}\,\widetilde{T}_{0I}(b)\Big]+\frac{\lambda}{\epsilon_{\mathcal{P}}}\widetilde{T}_{0R}(b)\,\widetilde{T}_{0I}(b)\,(e^{-\epsilon_{\mathcal{P}}\,\Delta\tau}-1)}{\Big[\Big(1-\frac{\lambda}{\epsilon_{\mathcal{P}}}\,\widetilde{T}_{0I}(b)\Big)e^{-\epsilon_{\mathcal{P}}\,\Delta \tau}+\frac{\lambda}{\epsilon_{\mathcal{P}}}\,\widetilde{T}_{0I}(b)\Big]^2+\frac{\lambda^2}{\epsilon_{\mathcal{P}}^2}\,\widetilde{T}_{0R}(b)^2\,\Big(1-e^{-\epsilon_{\mathcal{P}}\,\Delta\tau}\Big)^2}\nonumber \\
   &&= \int_0^{\infty} b\,db\,\frac{A\Big[\Big(e^{\beta\,b^2}-\frac{\lambda}{\epsilon_{\mathcal{P}}}B\Big)\,e^{-\epsilon_{\mathcal{P}}\,\Delta\tau}+\frac{\lambda}{\epsilon}_{\mathcal{P}}\,B\Big]-\frac{\lambda}{\epsilon_{\mathcal{P}}}AB\,(1-e^{-\epsilon_{\mathcal{P}}\,\Delta\tau})}{\Big[\Big(e^{\beta\,b^2}-\frac{\lambda}{\epsilon_{\mathcal{P}}}\,B\Big)e^{-\epsilon_{\mathcal{P}}\,\Delta \tau}+\frac{\lambda}{\epsilon_{\mathcal{P}}}\,B\Big]^2+\frac{\lambda^2}{\epsilon_{\mathcal{P}}^2}\,A^2\,\Big(1-e^{-\epsilon_{\mathcal{P}}\,\Delta\tau}\Big)^2}~\nonumber \\
   &&=\frac{\epsilon_{\mathcal{P}}\,e^{\epsilon_{\mathcal{P}}\,\Delta\tau}}{4\beta\lambda\,(e^{\epsilon_{\mathcal{P}}\,\Delta\tau}-1)}\Bigg(\pi+2\tan^{-1}\Big(\frac{\epsilon_{\mathcal{P}}-B\,\lambda+B\lambda\,e^{\epsilon_{\mathcal{P}}\Delta\tau}}{A\,\lambda(1-e^{\epsilon_{\mathcal{P}}\Delta\tau})}\Big)\Bigg)~,
   \label{eqa:Real_Analytic_Solution_Gaussian}
\end{eqnarray}
\begin{eqnarray}
   &&T_{I}(\tau,t=0)= \int_0^{\infty} b\,db\,\,\widetilde{T}_{0I}(\tau,b) \nonumber \\
   &&=\int_0^{\infty} b\,db\,\frac{\widetilde{T}_{0I}(b)\Big[\Big(1-\frac{\lambda}{\epsilon_{\mathcal{P}}}\widetilde{T}_{0I}(b)\Big)\,e^{-\epsilon_{\mathcal{P}}\,\Delta\tau}+\frac{\lambda}{\epsilon}_{\mathcal{P}}\,\widetilde{T}_{0I}(b)\Big]-\frac{\lambda}{\epsilon_{\mathcal{P}}}\widetilde{T}_{0R}(b)^2\,(e^{-\epsilon_{\mathcal{P}}\,\Delta\tau}-1)}{\Big[\Big(1-\frac{\lambda}{\epsilon_{\mathcal{P}}}\,\widetilde{T}_{0I}(b)\Big)e^{-\epsilon_{\mathcal{P}}\,\Delta \tau}+\frac{\lambda}{\epsilon_{\mathcal{P}}}\,\widetilde{T}_{0I}(b)\Big]^2+\frac{\lambda^2}{\epsilon_{\mathcal{P}}^2}\,\widetilde{T}_{0R}(b)^2\,\Big(1-e^{-\epsilon_{\mathcal{P}}\,\Delta\tau}\Big)^2}~\nonumber \\
   &&=\int_0^{\infty} b\,db\,\frac{B\Big[\Big(e^{\beta\,b^2}-\frac{\lambda}{\epsilon_{\mathcal{P}}}B\Big)\,e^{-\epsilon_{\mathcal{P}}\,\Delta\tau}+\frac{\lambda}{\epsilon}_{\mathcal{P}}\,B\Big]+\frac{\lambda}{\epsilon_{\mathcal{P}}}A^2\,(1-e^{-\epsilon_{\mathcal{P}}\,\Delta\tau})}{\Big[\Big(e^{\beta\,b^2}-\frac{\lambda}{\epsilon_{\mathcal{P}}}\,B\Big)e^{-\epsilon_{\mathcal{P}}\,\Delta \tau}+\frac{\lambda}{\epsilon_{\mathcal{P}}}\,B\Big]^2+\frac{\lambda^2}{\epsilon_{\mathcal{P}}^2}\,A^2\,\Big(1-e^{-\epsilon_{\mathcal{P}}\,\Delta\tau}\Big)^2}~ \nonumber \\
   &&=\frac{\epsilon_{\mathcal{P}}\,e^{\epsilon_{\mathcal{P}}\,\Delta\tau}}{4\,\beta\,\lambda(e^{\epsilon_{\mathcal{P}}\Delta\tau}-1)}\log\Bigg(1+\frac{2B\lambda}{\epsilon_{\mathcal{P}}}\Big(e^{\epsilon_{\mathcal{P}}\,\Delta\tau}-1\Big)+(A^2+B^2)\frac{\lambda^2}{\epsilon_{\mathcal{P}}^2}\Big(e^{\epsilon_{\mathcal{P}}\,\Delta\tau}-1\Big)^2\Bigg)~
   \label{eqa:Imaginary_Analytic_Solution_Gaussian}
\end{eqnarray}

\section*{Acknowledgments}

This paper is dedicated to my daughter, Josephine, whose curiosity and joy are a constant source of inspiration.
I am deeply grateful to Erasmo Ferreira and Takeshi Kodama, whose continuous encouragement, insightful comments, and long-standing support have played a crucial role in my development as a physicist. Even after my departure from the university, they have consistently taken the time to read and discuss my work, and their guidance has remained invaluable throughout the years.
I also thank Pedro C. Malta, Piotr Kotko, and Cyrille Marquet for carefully reading the manuscript, engaging in fruitful discussions, and providing important suggestions. I am grateful to Said Mazen for stimulating discussions on the mathematical aspects of the differential equations.
I gratefully acknowledge the AGH University of Science and Technology for its hospitality during the early stages of this project. Finally, I acknowledge the support of the National Science Centre in Poland, grant no. 2020/37/K/ST2/02665, and the Norwegian Financial Mechanism, during the period 2021–2023, when the initial ideas for this work were developed.




\end{document}